%% file: Manuscript.tex
\documentclass[journal,twoside]{IEEEtran}
\input{Preample}
\begin{document}
\title{Enabling High-Precision 5G mmWave-Based Positioning for Autonomous Vehicles in Dense Urban Environments}

\input{Authors}
\maketitle
\bstctlcite{IEEEexample:BSTcontrol} 

\begin{abstract}
5G-based mmWave wireless positioning has emerged as a promising solution for autonomous vehicle (AV) positioning in recent years. Previous studies have highlighted the benefits of fusing a line-of-sight (LoS) 5G positioning solution with an Inertial Navigation System (INS) for an improved positioning solution. However, the highly dynamic environment of urban areas, where AVs are expected to operate, poses a challenge, as non-line-of-sight (NLoS) communication can deteriorate the 5G mmWave positioning solution and lead to erroneous corrections to the INS. To address this challenge, we exploit 5G multipath and LoS signals to improve positioning performance in dense urban environments. In addition, we integrate the proposed 5G-based positioning with low-cost onboard motion sensors (OBMS). Moreover, the integration is realized using an unscented Kalman filter (UKF) as an alternative to the widely utilized EKF as a fusion engine to avoid ignoring the higher-order and non-linear terms of the dynamic system model. We also introduce techniques to evaluate the quality of each LoS and multipath measurement prior to incorporation into the filter's correction stage. To validate the proposed methodologies, we performed two test trajectories in the dense urban environment of downtown Toronto, Canada. For each trajectory, quasi-real 5G measurements were collected using a ray-tracing tool incorporating 3D map scans of real-world buildings, allowing for realistic multipath scenarios. For the same trajectories, real OBMS data were collected from two-different low-cost IMUs. Our integrated positioning solution was capable of maintaining a level of accuracy below $30$ cm for approximately $97\%$ of the time, which is superior to the accuracy level achieved when multipath signals are not considered, which is only around $91\%$ of the time.
\end{abstract}

\def\abstractname{Note to Practitioners}
\begin{abstract}
Autonomous vehicles are gaining popularity but require highly accurate positioning to operate safely. Achieving decimeter-level accuracy for at least $95\%$ of the time is challenging in dense urban environments where GPS signals may be blocked. This paper proposes using 5G wireless networks to provide high-precision positioning services to address this issue, as 5G base stations are expected to be densely deployed in urban areas. However, maintaining a line-of-sight (LoS) communication with 5G base stations may not always be possible in dense urban areas due to the multi-path from surrounding buildings. Therefore, we suggest fusing LoS measurements with non-line-of-sight (NLoS) measurements to improve positioning accuracy in challenging urban environments. To guarantee seamless positioning even in scenarios involving 5G signal outages, we also incorporate onboard motion sensors like accelerometers, gyroscopes, and odometers to ensure that the autonomous vehicle's positioning remains accurate and reliable even in all challenging urban environments.
\end{abstract}


\section{Introduction}
\IEEEPARstart{A}{utonomous} vehicles (AVs) are becoming increasingly important in the transportation industry as they have the potential to greatly improve safety, reduce traffic congestion, and provide more efficient transportation. However, AVs rely heavily on absolute positioning systems to navigate and operate safely \cite{Reid_2019}. While global navigation satellite systems (GNSS) are often used for this purpose, they can be unreliable in urban areas due to the high-rise buildings that can block or reflect GNSS signals \cite{NovAtel}. This can make it difficult for AVs to accurately determine their location and orientation, which is essential for safe operation. On the other hand, onboard motion sensors (OBMS), like accelerometers, gyroscopes, and odometers, do not suffer from the aforementioned problems as they are self-contained. OBMS measurements can be processed using a dead reckoning algorithm like the inertial navigation system (INS) to compute the vehicle's position, velocity and attitude. INS has the advantage of providing the positioning solution at a high data rate. However, the inherent errors of the OBMS may result in growing position errors if they work in standalone mode, which can be resolved by pairing it with other reliable positioning technologies of superior accuracy (such as 5G wireless positioning) to estimate and reset the INS errors \cite{ProfBook}.

\IEEEpubidadjcol  

Recently, 5G NR mmWave has been explored as a potential positioning technology for AVs \cite{REF22,Would5G}. The high-frequency band of the 5G wireless spectrum provides a high bandwidth of $400$ MHz, allowing for accurate time-based measurements like time of arrival (ToA), round-trip time (RTT), and time difference of arrival (TDoA), as well as the ability to resolve multipath components (MPC) in the time domain. Massive multi-input-multi-output (MIMO) capabilities of mmWave allow for accurate angle-based measurements such as angle of arrival (AoA) and angle of departure (AoD). 5G mmWave also features low latency communications, making it ideal for supporting the real-time decision-making and navigation of AVs. By leveraging the unique propagation characteristics of mmWave signals, it is possible to achieve decimeter-level positioning accuracy, which is essential for AVs' safe and reliable operation. Finally, 5G small cells are expected to be densely deployed every $200$ m to $500$ m, which means that vehicular systems will endure a higher chance of LoS connectivity with the deployed gNBs \cite{Survey9}.

Despite the higher line-of-sight (LoS) probability associated with 5G technology in comparison to LTE, the user equipment (UE) may still encounter non-line-of-sight (NLoS) communication. This is attributed to the dynamic nature of urban environments, where various physical obstacles, such as buildings, trees, pedestrians, buses, and trucks, can impede the 5G signal. Directly using NLoS signals for positioning using LoS-based algorithms will significantly bias the positioning solution. The literature offers several approaches to address the NLoS issue. One line of research focuses on techniques that minimize the effect of NLoS links on positioning accuracy \cite{nlosmit1,nlosmit2,nlosmit3,nlosmit4}. In contrast, others aim to detect and discard NLoS signals to avoid positioning errors caused by multipath \cite{nlosdiscard1,nlosdiscard2}. However, recent work explores multipath rays as an additional source of positioning information during 5G outages \cite{nlosuse1,nlosuse2}. This paper introduces a new, high-precision accurate positioning solution that combines LoS, multipath 5G mmWave-based signals, and OBMS to offer an uninterrupted positioning solution at a high data rate, suitable for AV operation in dense urban areas. To the best of our knowledge, no existing literature fuses multipath signals with OBMS. To expand on this, we propose a measurement selection scheme to evaluate each multipath measurement before fusion. Additionally, we suggest using the unscented Kalman filter (UKF) as an alternative to the commonly used extended Kalman filter (EKF) to avoid the errors associated with the linearization of the dynamic and measurement system models, as described in \cite{decentralized} and as demonstrated in our analysis later in this paper. Through rigorous testing, we demonstrate that the proposed solution achieves exceptional performance over two distinct trajectories with varying dynamics and 5G outage probabilities and with different suites of low-cost OBMS.

The contributions of this paper are as follows:
\begin{enumerate}
  \item We present an enhanced positioning solution based on the loosely-coupled (LC) integration of 5G LoS and multipath signals with OBMS utilizing a UKF.
  \item We employ a measurement exclusion scheme that relies on the UE and BS propagation link.
  \item We propose an additional validation stage for 5G NLoS measurements using constraints derived from odometer measurements. 
  \item For validation, we conducted two road test trajectories in downtown Toronto (Ontario, Canada) involving  actual OBMS measurements collected from sensors mounted inside the test vehicle and integrated with a quasi-real 5G-based mmWave observables generated by the $S_5G$ simulation software, which accurately emulates the complex urban environment of Toronto's downtown area, where the road tests were conducted.
\end{enumerate}

The paper is structured as follows: Section II presents a literature review. Section III outlines the system model, covering the foundations of 5G and INS measurements and various Kalman filter implementations. Section IV proposes a 5G/OBMS LC integration approach using a UKF. Section V provides information about the experimental and road test setup. Section VI presents the results and discussions. Finally, Section VII concludes the paper.

\section{Literature Review}
Very limited works have integrated 5G measurements with OBMS \cite{LR1,LR2,LR3}. The work in \cite{LR1} utilized federated filtering to integrate INS/5G/GPS/LEO by means of sub-filters reporting to a central filter. In their 5G/INS sub-filter, they fuse 5G pseudo-range measurements with INS measurements by means of tight coupling (TC) utilizing an EKF. Such integration will yield high linearization error as the transition and observation models are non-linear. Furthermore, excluding angle-based 5G measurements can also constrain positioning accuracy and mandate the UE to establish connections with at least three BSs concurrently to obtain a precise 3D positioning solution. Relying on trilateration assuming access to three or more base stations may not be possible in dense urban areas and would result in a severe multipath effect that deteriorates the positioning accuracy. In reference to \cite{LR2}, the authors fused INS with 5G ToA and AoA. During the prediction stage, they rely on accelerometer readings to estimate velocity and position by incorporating a constant acceleration model. However, it is worth noting that such a model may be considered unusual given that INS mechanization techniques are already established in the literature and could offer more reliable computation of position and velocity at higher data rates without imposing limitations on the vehicle dynamics (e.g. constant acceleration model). In addition, using an EKF for filtering in the presence of non-linear transition and measurement models may result in sub-optimal performance. Lastly, their IMU measurements are simulated, making it hard to generalize or compare with other positioning solutions. For instance, simulated IMU data may not account for the effects of external factors such as temperature changes, magnetic interference, and mechanical vibrations. As a result, the performance of the positioning solution based on simulated data may not generalize well to real-world scenarios, which is addressed in this paper. Finally, the approach proposed in \cite{LR3} suggests the use of a constant acceleration model for the prediction stage and 5G ToA, AoA, and IMU accelerations in the $x$ and $y$ directions for corrections. The EKF is utilized for the final integration, where the UE position, velocity, and acceleration are considered system states. However, this method does not consider estimating the azimuth (heading) angle, an essential variable for the navigation solution in real-life operations. Additionally, the direct use of ToA and AoA measurements in the measurements vector leads to linearization errors, as mentioned earlier.

\section{System Model}
\subsection{5G System Model}
We take into account a down-link 3D positioning scenario with a single base station (BS) and a single UE with positions $\boldsymbol{p}_{b_{3D}} = \begin{bmatrix}x_b& y_b& z_b\end{bmatrix}^T$ and $\boldsymbol{p}_{3D} = \begin{bmatrix}x& y& z\end{bmatrix}^T$ respectively. We assume that the position of the BS is known and that the BS and the UE are oriented in a given manner. We use the channel parameters like AoA, denoted by $\beta$, AoD, denoted by $\alpha$, and ToA, denoted by $\tau$, for each path. ToA can be used to compute the range between the BS and the UE through the following formula:

\begin{equation}
    \tau = \frac{d_{3D}}{c},
\end{equation}


\noindent where $d_{3D}$ is the total propagation distance, and $c$ is the speed of light. The use of ToA requires tight synchronization between the UE and the BS. Else, time bias will afflict the measurements, causing positioning errors. RTT and TDoA measurements, on the other hand, do not require synchronization between the UE and the BS. In this paper, RTT measurements are utilized. To compute the range between the BS and the UE, the RTT measurement should be first divided by two to account for the total distance travelled. This work assumes that the UE will only be connected to the nearest BS. To determine the type of communication link between the BS and the UE, an NLoS detection technique based on range comparisons between RTT and RSS measurements, proposed in \cite{NLOS}, will be used.

\subsubsection{5G LoS Positioning} \label{sec: 5G Hybrid}
The AoD and RTT information obtained from a single BS are used to determine the 3D position of the UE. The AoD provides the direction of the signal sent to the UE, while the RTT information can be used to calculate the distance from the BS to the UE. These measurements can then be used to determine the 3D position of the UE as seen in (\ref{5G_Hybrid}).

\begin{equation}\label{5G_Hybrid}
\boldsymbol{p}_{3D}=\boldsymbol{p}_{b_{3D}} + d_{3D}
\begin{bmatrix}
\sin\alpha \cos\phi\\
\cos\alpha \cos\phi\\
\sin(\phi)
\end{bmatrix}
\end{equation}

\noindent Where $d_{3D}$ is the measured 3D distance between the BS and the UE, and $\alpha$ and $\phi$ are the estimated horizontal and elevation AoD angles, respectively. If a constant height assumption can be made about the UE, then the 3D positioning equation can be simplified to estimate the 2D position of the UE as seen in (\ref{P_2D}).

\begin{equation}\label{P_2D}
\boldsymbol{p} = \boldsymbol{p}_b + d
\begin{bmatrix}
\sin\alpha \\
\cos\alpha
\end{bmatrix}
\end{equation}

\noindent Where $\boldsymbol{p}$ is the estimated 2D position of the UE, $\boldsymbol{p}_b$ is the 2D position of the BS, and $d$ is the 2D distance from the BS to the UE.

\subsubsection{5G Multipath Positioning} \label{sec: MPP}
\begin{figure}[h]
	\centering
	\includegraphics[width=\columnwidth]{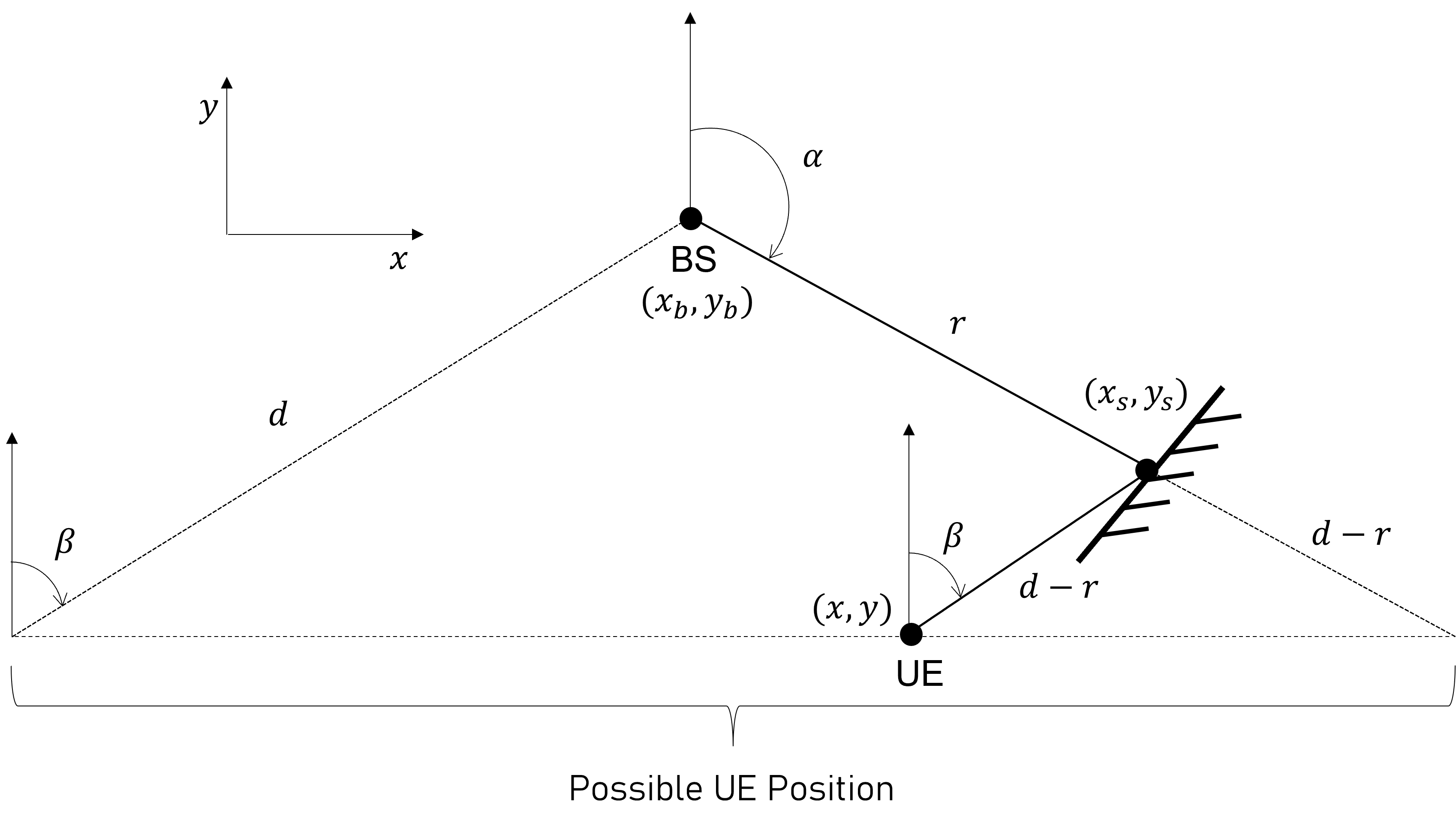}
	\DeclareGraphicsExtensions.
	\caption{System model of a single-bounce reflection scenario.}
	\label{fig:MPP}
\end{figure}

The SBR-based positioning scheme introduced in \cite{SB0} is utilized. The algorithm determines the segment of possible UE position by utilizing AoD, AoA, and the distance $d$ of the strongest propagation path, as depicted in Fig. \ref{fig:MPP}.

\noindent The figure displays the system model of a single-bounce reflection scenario. The scatterer's coordinates, $\boldsymbol{p_{s}}=\begin{bmatrix}x_s&y_s\end{bmatrix}^T$, and the UE's coordinates, $\boldsymbol{p}$, are calculated as seen in (\ref{Scatt}) and (\ref{Scatt2}).

\begin{equation}\label{Scatt}
\boldsymbol{p_{s}} = \boldsymbol{p_{b}} + r
\begin{bmatrix}
\sin\beta \\
\cos\beta
\end{bmatrix}, \qquad r\in(0,d)
\end{equation}

\begin{equation} \label{Scatt2}
\boldsymbol{p} = \boldsymbol{p_{s}} - (d - r)
\begin{bmatrix}
\sin\alpha \\
\cos\alpha
\end{bmatrix} , \qquad r\in(0,d)
\end{equation}

\noindent Where $r$ is the distance between the BS and the scatterer. The possible position of the UE can be represented by a straight-line equation as seen in (\ref{Straightline}).
\begin{equation}\label{Straightline}
y=k(\alpha,\beta)x+b(\alpha,\beta,d)
\end{equation}
\noindent Where,
\begin{equation}
k(\alpha,\beta) = \frac{\cos \alpha + \cos \beta}{\sin \alpha + \sin \beta},
\end{equation}
\noindent and,
\begin{equation}
b(\alpha,\beta,d) = -k(\alpha,\beta)(x_{b} - d\sin\alpha) + y_{b} - d\cos\alpha.
\end{equation}

Accordingly, the position of the UE can be determined by finding the intersection between two lines of two propagation paths, if available. This work uses an order-of-reflection identification  (OoRI) technique to filter out higher-order reflections. The technique is based on ensemble learning and relies on 5G channel parameters, such as AoA, AoD, ToA, and RSS \cite{OoRI}.

\section{OBMS System Model}
\subsection{INS Measurables}
A typical INS comprises an IMU unit consisting of three accelerometers and three gyroscopes. These sensors measure, along three mutually orthogonal directions,  the accelerations $f_x, f_y$ and $f_z$ and angular rates $\omega_x, \omega_y$, and $\omega_z$ of a moving body in a 3D space. Such measurements are often used for dead-reckoning positioning, which involves estimating the current position of a moving body based on its previous position, velocity, and orientation states. By integrating the specific forces and angular rate measurements from an IMU over time, it is possible to estimate the displacement and orientation of the object relative to its starting position. To achieve this, the accelerometer and gyroscope readings must be converted from the body frame, also known as the b-frame, to a global Earth-fixed coordinate frame. A local-level frame, also known as the l-frame, is frequently used, as seen in (\ref{fl}). Such transformation utilizes the $\boldsymbol{R}_b^l$ rotation matrix as defined in \cite{ProfBook} that transforms the measurement from the body frame (b) to the local navigation frame (l).

\begin{equation}\label{fl}
    \begin{split}
        \boldsymbol{f}_l&=\boldsymbol{R}_b^l \boldsymbol{f}_b\\
        \boldsymbol{\omega}_l&=\boldsymbol{R}_b^l \boldsymbol{\omega}_b,\\
    \end{split}
\end{equation}
\begin{figure*}[h]
\normalsize
\begin{equation}\label{RBL}
\boldsymbol{R}_b^l=\begin{bmatrix}
 \cos{a}\cos{r} + \sin{a}\sin{p}\sin{r} & \sin{a}\cos{p} & \cos{a}\sin{r} - \sin{a}\sin{p}\cos{r}\\
-\sin{a}\cos{r} + \cos{a}\sin{p}\sin{r} & \cos{a}\cos{p} & -\sin{a}\sin{r} - \cos{a}\sin{p}\cos{r}\\
-\cos{p}\sin{r} & \sin{p} & \cos{p}\cos{r}
\end{bmatrix}
\end{equation}
\end{figure*}

Among the errors associated with the OBMS are the sensors' noise and bias. Sensor noise is the random fluctuations in the sensor output due to the inherent sensor design and possibly the surrounding environment. The bias has two components. The first is a deterministic offset that can be removed by calibration. The second is the bias drift which is stochastic in nature that changes over time, even when no external forces or rotation are present. Sensor fusion and calibration are frequently used to combine data from multiple sensors to estimate and correct such errors \cite{IMUErrors}.

\subsection{Odometers}
A wheel odometer that provides the vehicle's forward speed in the b-frame is utilized. However, since our states are in the l-frame, we need to transform the odometer velocities in the b-frame, denoted as $v_b = \begin{bmatrix}0& v_{Odo}& 0\end{bmatrix}^T$, using the second column of the rotation matrix $\boldsymbol{R_b^l}$ as shown in (\ref{Vel}).

\begin{equation} \label{Vel}
\begin{bmatrix}
v_e\\
v_n\\
v_u
\end{bmatrix} = 
\begin{bmatrix}
\sin{a}\cos{p}\\
\cos{a}\cos{p}\\
\sin{p}
\end{bmatrix} v_{Odo}
\end{equation}

\section{5G-OBMS Integration Scheme} \label{Methodology}
\begin{figure*}[h]
	\centering
	\includegraphics[width=500pt]{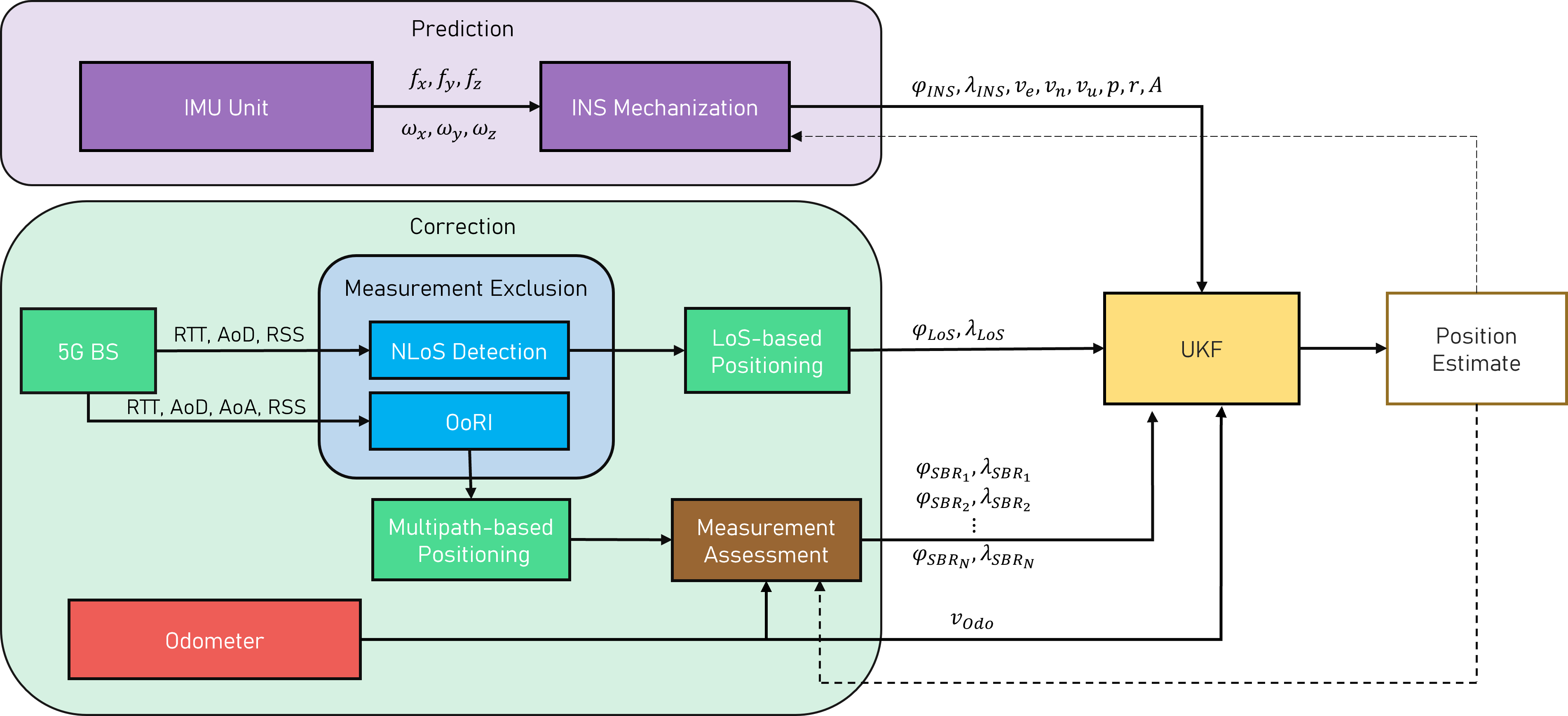}
	\DeclareGraphicsExtensions.
	\caption{Block diagram of the proposed integrated positioning system.}
	\label{method}
\end{figure*}

Within this section, we propose the utilization of a UKF \cite{UKF} to incorporate 5G measurements, derived from both LoS and multipath sources, with OBMS in a loosely-coupled (LC) manner. This method of integration fuses independent position estimates obtained from OBMS, 5G LoS, and 5G NLoS measurements, in contrast to tightly-coupled (TC) integration that directly fuses raw 5G and OBMS measurements. The latter approach leads to high linearization errors, as discussed in \cite{decentralized}. The states vector $\boldsymbol{x}$, the state transition model $f(\boldsymbol{x}$, $\boldsymbol{u}$), and the process covariance matrix $\boldsymbol{Q}$ of the proposed method will be displayed first. Then, the proposed measurement vector $\boldsymbol{z}$, together with the measurement model $h(\boldsymbol{x})$ and the noise covariance matrix $\boldsymbol{R}$ are presented next. Finally, we showcase the proposed measurement assessment strategy based on the vehicle's movement constraints. The overall block diagram of the proposed system is shown in Fig.\ref{method}.

\subsection{States and States Transition Model}
The proposed method estimates the positioning states in the geodetic reference frame, namely, latitude~$\varphi$, longitude~$\lambda$, and altitude~$h$. In addition to the positioning states, the velocity component along the east, north, and up (ENU) directions are also estimated, denoted by $v_e$, $v_n$, and $v_u$, respectively. Lastly, attitude components comprise the pitch $p$, roll $r$, and azimuth $A$ angles. The aforementioned states are collectively referred to as the PVA states and are shown in (\ref{States}).

\begin{equation}\label{States}
    \boldsymbol{x}_{PVA}=\begin{bmatrix}
        \varphi&
        \lambda&
        h&
        v_e&v_n&v_u&
        p&r&A& 
    \end{bmatrix}^T
\end{equation}

The aforementioned states are momentarily augmented with the system inputs, represented by the vector $\boldsymbol{u} = \begin{bmatrix} f_x& f_y& f_z& \omega_x& \omega_y& \omega_z \end{bmatrix}$, which encompasses acceleration and angular velocity measurements. This preliminary stage precedes the generation of sigma points, with the objective of producing a uniform set of $2n+1$ sigma points for INS measurements. This facilitates the capacity of the Unscented Kalman Filter (UKF) to characterize the impact of the inputs on the system state, thereby improving the accuracy of the system's actual state estimation.

The proposed transition model $f(\boldsymbol{x},\boldsymbol{u})$ is governed by the INS mechanization process. INS mechanization is the process of computing the navigation PVA states from the raw inertial measurements. The mathematical representation of INS mechanization in the l-frame can be summarized in Eqs. (\ref{P}-\ref{A}):

\begin{equation}\label{P}
    \begin{bmatrix}
        \dot{\varphi}\\
        \dot{\lambda}\\
        \dot{h}
    \end{bmatrix}
    = 
    \begin{bmatrix}
        0& \frac{1}{R_M + h}& 0\\
        \frac{1}{(R_N + h)cos\varphi}& 0& 0&\\
        0& 0& 1&
    \end{bmatrix}
     \begin{bmatrix}v_e\\ v_n\\ v_u\end{bmatrix}
\end{equation}

Eq. (\ref{P}) demonstrates the relationship between the geodetic coordinates denoted by $\dot{\varphi}, \dot{\lambda}, \text{and} ~\dot{h}$ and the velocities along the l-frame denoted by $v_e, v_n, \text{and}~ v_u$. $R_N$ is the radius of curvature in the Prime Vertical, and $R_M$ is the radius of curvature in the Meridian. Eq. (\ref{V}) represents the velocity mechanization in the l-frame.

\begin{equation} \label{V}
    \boldsymbol{\dot{v}}^l = \boldsymbol{R^l_b} \boldsymbol{f}^b - (2\boldsymbol{\Omega}^l_{ie} + \boldsymbol{\Omega}^l_{el})\boldsymbol{v}^l + \boldsymbol{g}^l,
\end{equation}

\noindent where $\boldsymbol{\dot{v}}^l$ is the kinematic acceleration in the l-frame. The components $2\boldsymbol{\Omega}^l_{ie} \cdot \boldsymbol{v}^l$, and $\boldsymbol{\Omega}^l_{el}\cdot \boldsymbol{v}^l$ denote the acceleration observed in the l-frame with respect to the Earth frame (e-frame), and the Coriolis acceleration, respectively. In particular, $\boldsymbol{\Omega}^l_{ie}$ is the skew-symmetric matrix of $\boldsymbol{\omega}_{ie}^l$, which is a vector that represents the Earth's rotation rate in the l-frame as seen in (\ref{wiel}).

\begin{equation} \label{wiel}
    \boldsymbol{\omega}_{ie}^l = [0~\omega^e cos\varphi~\omega^e sin\varphi]^T
\end{equation}

\noindent $\boldsymbol{\Omega}^l_{el}$ is a skew-symmetric matrix of $\boldsymbol{\omega}_{el}^l$ representing the rotation rate of the l-frame relative to the e-frame and expressed in the l-frame as seen in (\ref{well}).

\begin{equation}\label{well}
    \boldsymbol{\omega}_{el}^l = \left[\frac{-v_n}{R_M+h}~\frac{v_e}{R_N+h}~\frac{v_e tan\varphi}{R_N+h} \right]^T
\end{equation}

\noindent Furthermore, $\boldsymbol{g}^l = \begin{bmatrix}0&0&-g\end{bmatrix}^T$ is the gravity vector. Lastly, solving the time derivative equation of the transformation matrix $\boldsymbol{R}_l^b$ yields the attitude (orientation) of the moving body as seen in (\ref{A}).

\begin{equation} \label{A}
    \boldsymbol{\dot{R}}^l_b = \boldsymbol{R^l_b} (\boldsymbol{\Omega}^b_{ib} + \boldsymbol{\Omega}^b_{il})
\end{equation}

\noindent Where $\boldsymbol{\Omega}^b_{ib}$ is a skew-symmetric matrix of $\boldsymbol{\omega}^b_{ib}$ representing the gyroscope measurements that encode the rotation rate of the b-frame relative to the earth-centred-inertial (ECI) frame and expressed in the b-frame. The $\boldsymbol{\Omega}^b_{il}$ is the skew-symmetric matrix of $\boldsymbol{\omega}^b_{il}$ representing the rotation rate of the l-frame relative to the inertial frame expressed in the b-frame. It can be computed by adding $\boldsymbol{\omega}^l_{ie}$ and $\boldsymbol{\omega}^l_{el}$ as seen in (\ref{wilb}).

\begin{equation}\label{wilb}
    \boldsymbol{\omega}^b_{il} = \boldsymbol{R}_l^b \cdot(\boldsymbol{\omega}^l_{ie} + \boldsymbol{\omega}^l_{el})
\end{equation}

\noindent The summary of the transition system model $f(\boldsymbol{x}_{k-1}^+,\boldsymbol{u}_k)$ can be seen in (\ref{PVA}).

\begin{equation} \label{PVA}
    \begin{bmatrix}
        \dot{\boldsymbol{r}^l}\\
        \dot{\boldsymbol{v}^l}\\
        \dot{\boldsymbol{R}^l_b}\\
    \end{bmatrix} =
    \begin{bmatrix}
        \boldsymbol{D}^{-1}\boldsymbol{v}^l\\
        \boldsymbol{R^l_b} \boldsymbol{f}^b - (2\boldsymbol{\Omega}^l_{ie} + \boldsymbol{\Omega}^l_{el})\boldsymbol{v}^l + \boldsymbol{g}^l\\
        \boldsymbol{R^l_b} (\boldsymbol{\Omega}^b_{ib} + \boldsymbol{\Omega}^b_{il})
    \end{bmatrix}
\end{equation}

\begin{figure}[h]
	\centering
	\includegraphics[width=\columnwidth]{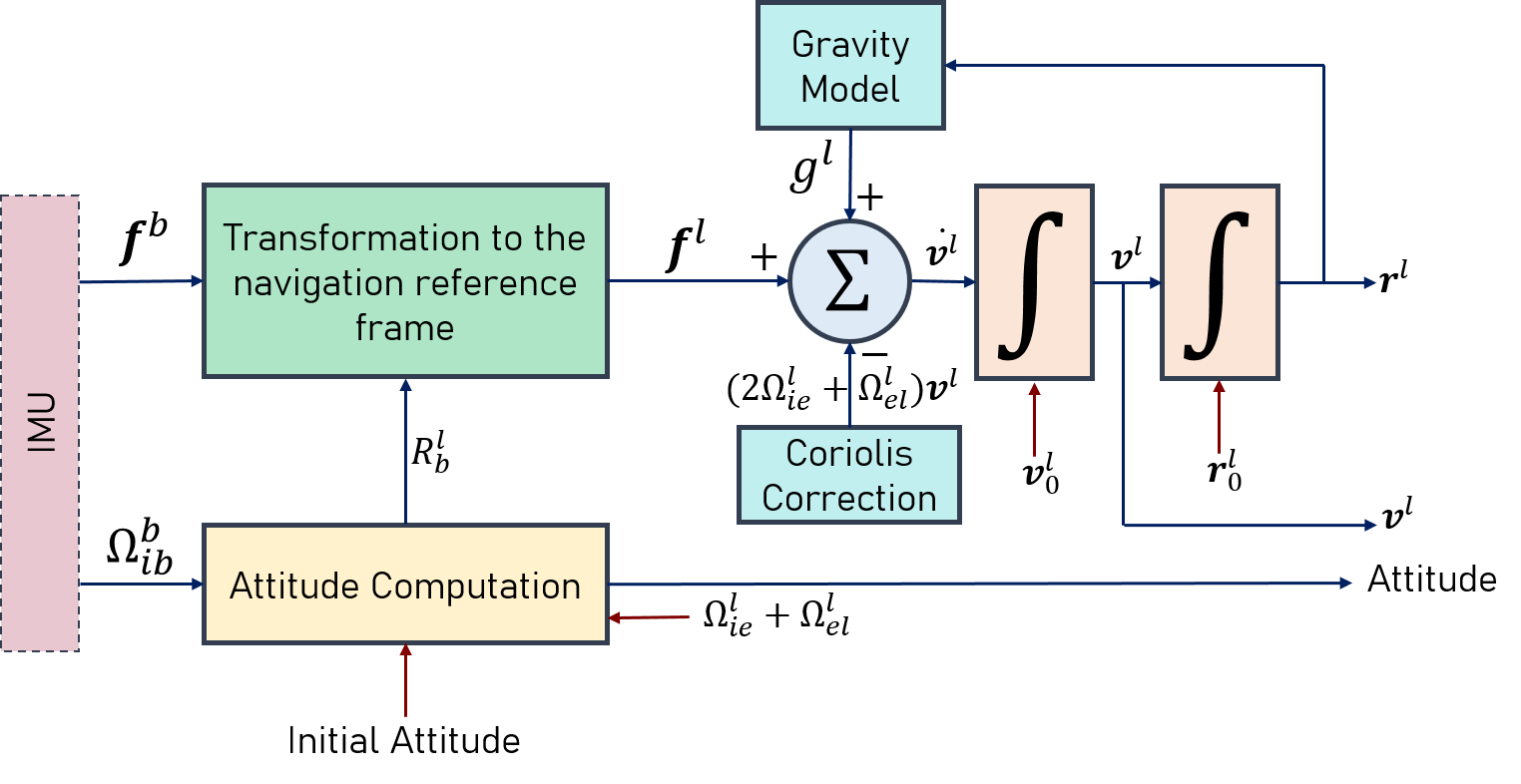}
	\DeclareGraphicsExtensions.
	\caption{Detailed INS mechanization block diagram}
	\label{ins detailed}
\end{figure}

Where $\dot{\boldsymbol{r}^l}$ is the time rate of change of the three position components, $\varphi, \lambda$, and $h$, and $\boldsymbol{D}^{-1}$ is defined as follows:

\begin{equation}
    \boldsymbol{D}^{-1} = 
        \begin{bmatrix}
        0& \frac{1}{R_M + h}& 0\\
        \frac{1}{(R_N + h)cos\varphi}& 0& 0&\\
        0& 0& 1&
    \end{bmatrix}
\end{equation}
Fig. \ref{ins detailed} presents the detailed block diagram of INS mechanization.

\subsubsection{Quaternions}
The parameterization of the rotation matrix  $\boldsymbol{R^l_b}$ is necessary to solve the mechanization equations. The use of quaternions is a widely adopted technique in many fields of study, owing to its numerous advantageous features \cite{ProfBook}. For instance, the quaternion solution does not suffer from the problem of gimbal lock, which can be a major issue when using other rotation representations, such as Euler angles. A Gimbal lock occurs when two or more of the rotation axes align, resulting in a loss of one degree of freedom and making certain rotations impossible to represent. Additionally, quaternion computations are relatively simple to perform. Quaternions are composed of four components: a scalar part and a vector part. The scalar part is a real number, while the vector part is a three-dimensional vector, and is defined as follows:

\begin{equation}\label{Quat}
\boldsymbol{q} = \begin{bmatrix}
\frac{0.25*(r_{32}-r_{23})}{q_4} \\
\frac{0.25*(r_{13}-r_{31})}{q_4} \\
\frac{0.25*(r_{21}-r_{12})}{q_4} \\
0.5*\sqrt{1+r_{11}+r_{22}+r_{33}}
\end{bmatrix}
\end{equation}


Where the notation $r_{12}$ indicates the first row and second column element of the rotation matrix $\boldsymbol{R_b^l}$, and $q_4$ denotes the fourth element of the quaternion vector $\boldsymbol{q}$. The components of a quaternion are typically subject to certain constraints. Specifically, in some contexts, the components of a quaternion may be required to have a norm or magnitude of 1 as seen in (\ref{one}). This norm constraint ensures that the quaternion represents a rotation, and it is often referred to as the unit quaternion constraint.

\begin{equation}\label{one}
    q_1^2 + q_2^2 + q_3^2 + q_4^2 = 1
\end{equation}

The aforementioned equivalence might not hold true due to computational errors. To compensate for this, the quaternion parameters vector Following each computational step, $\boldsymbol{q}$ needs to be updated as follows:

\begin{equation}
    \hat{\boldsymbol{q}} = \frac{\boldsymbol{q}}{\sqrt{1-\Delta}}\cong \boldsymbol{q}\left(1+\frac{\Delta}{2}\right)
\end{equation}

\noindent where,

\begin{equation}
    \Delta = 1- (q_1^2 + q_2^2 + q_3^2 + q_4^2)
\end{equation}

In order to predict quaternion components $\boldsymbol{q}_{k+1}$ based on $\boldsymbol{q}_k$, the following formula is used:

\begin{equation}
    \boldsymbol{q}_{k+1}=\boldsymbol{q}_k+\left(\frac{1}{2} \Omega_{il}^b\left(\omega_k\right) \boldsymbol{q}_k\right) \Delta t,
\end{equation}

\noindent where $\omega_k$ is the angular velocities of body rotations. The following direct relationship can be used to find the rotation matrix $\boldsymbol{R}_b^l$ once the quaternion parameters have been established as seen in (\ref{RBL-Quat}).

\begin{figure*}[h]
\normalsize
\begin{equation} \label{RBL-Quat}
\boldsymbol{R}_b^l =
\begin{bmatrix}
\boldsymbol{q}_{(1)}^2-\boldsymbol{q}_{(2)}^2-\boldsymbol{q}_{(3)}^2+\boldsymbol{q}_{(4)}^2 & 2\boldsymbol{q}_{(1)}\boldsymbol{q}_{(2)}+2\boldsymbol{q}_{(3)}\boldsymbol{q}_{(4)} & 2\boldsymbol{q}_{(1)}\boldsymbol{q}_{(3)}-2\boldsymbol{q}_{(2)}\boldsymbol{q}_{(4)} \\
2\boldsymbol{q}_{(1)}\boldsymbol{q}_{(2)}-2\boldsymbol{q}_{(3)}\boldsymbol{q}_{(4)} & -\boldsymbol{q}_{(1)}^2+\boldsymbol{q}_{(2)}^2-\boldsymbol{q}_{(3)}^2+\boldsymbol{q}_{(4)}^2 & 2\boldsymbol{q}_{(2)}\boldsymbol{q}_{(3)}+2\boldsymbol{q}_{(1)}\boldsymbol{q}_{(4)} \\
2\boldsymbol{q}_{(1)}\boldsymbol{q}_{(3)}+2\boldsymbol{q}_{(2)}\boldsymbol{q}_{(4)} & 2\boldsymbol{q}_{(2)}\boldsymbol{q}_{(3)}-2\boldsymbol{q}_{(1)}\boldsymbol{q}_{(4)} & -\boldsymbol{q}_{(1)}^2-\boldsymbol{q}_{(2)}^2+\boldsymbol{q}_{(3)}^2+\boldsymbol{q}_{(4)}^2
\end{bmatrix}
\end{equation}
\end{figure*}


According to the rotation matrix $\boldsymbol{R}_b^l$ defined in (\ref{RBL}), the attitude angles can be computed using the newly computed matrix utilizing the following relationships:

\begin{equation}
p=tan^{-1} \left(\frac{r_{32}}{\sqrt{r_{12}^2+r_{22}^2}}\right)
\end{equation}
\begin{equation}
r=-tan^{-1} \left(\frac{r_{31}}{r_{33}}\right)
\end{equation}

\begin{equation}
   A=tan^{-1} \left(\frac{r_{12}}{r_{22}}\right) 
\end{equation}

\subsubsection{Process Covariance Matrix}
In contrast to prior works, we adopt a diagonal process noise covariance matrix $\boldsymbol{Q}$ representing the noises of the accelerometers and gyroscopes only, rather than encompassing all states noises, as seen in (\ref{Q}). 

\begin{equation} \label{Q}
    \boldsymbol{Q} = diag([\sigma_{\omega_x}^2~\sigma_{\omega_y}^2~\sigma_{\omega_z}^2~\sigma_{f_x}^2~\sigma_{f_y}^2~\sigma_{f_z}^2])
\end{equation}

\noindent Where $\sigma_{\omega_x}^2$, $\sigma_{\omega_y}^2$, and $\sigma_{\omega_z}^2$ are gyroscopic noises and $\sigma_{f_x}^2$,$\sigma_{f_y}^2$, and $\sigma_{f_z}^2$ are accelerometer noises. All of which are additive white Gaussian noise (AWGN). The design of the process covariance matrix in this way makes it easily tunable as the uncertainties of the system states are influenced by the uncertainties of system inputs which are propagated to the states through the transition model. In order to produce the sigma points, it becomes necessary to augment the P and Q matrices to account for sensor noises.

\subsection{Measurements and Measurements Model}
\subsubsection{Measurements}
In the proposed method, the measurement vector $\boldsymbol{z}$ comprises the 3D position of the UE from both LoS and NLoS measurements. Additionally, it consists of the vehicle velocity with respect to the l-frame as acquired from a wheel odometer, as shown in (\ref{z}).

\begin{equation}\label{z}
    \boldsymbol{z}= \begin{bmatrix}
        \boldsymbol{\varphi}_{5G} & \boldsymbol{\lambda}_{5G} & \boldsymbol{h}_{5G} &  v_{e_{Odo}} & v_{n_{Odo}} & v_{u_{Odo}}
    \end{bmatrix}^T
\end{equation}

\noindent Where $\boldsymbol{\varphi}_{5G}$, $\boldsymbol{\lambda}_{5G}$, and $\boldsymbol{h}_{5G}$ are the 3D UE position measurements provided by 5G LoS and NLoS signals; and $v_{e_{Odo}}$, $v_{n_{Odo}}$, and $v_{u_{Odo}}$ are the vehicle velocity measurements provided by the odometer in the l-frame. 

\subsubsection{Measurement Exclusion}
It is crucial to highlight that the measurement vector $\boldsymbol{z}$ is subject to dynamic changes depending on the availability of LoS signals and SBRs. Prior to any positioning estimation, a measurement exclusion process is performed to filter out NLoS signals, allowing only LoS signals to be utilized by the LoS-based positioning module. This process follows our previous work described in \cite{NLOS}. The approach relies on the distinction in distance computation between the UE and the BS through the utilization of time-based and received signal strength-based calculations. On the other hand, when multipath signals are used for positioning, channel parameters are passed to an OoRI module, which filters out higher-order reflections by allowing only single-bounce reflections to be passed on to the multipath positioning module. The functioning of this OoRI module is presented in \cite{OoRI}. The machine learning model was trained on a dataset comprising $3.6$ million observations, which consisted of 5G channel parameters such as ToA, AoA, AoD, and Received Signal Strength (RSS). The training process involved using ensemble learning, where a total of $14$ decision tree learners were trained. Upon completion of the training, the model attained a classification accuracy of $99.8\%$.

\subsubsection{Measurement Assessment}
Given that the proposed OoRI model is based on machine learning, it is essential to address the issue of misclassified SBRs, which could result in substantial errors in the computed position if they are passed to the multipath positioning module. Hence, position computations resulting from multipath positioning undergo a second stage of validation, which is contingent upon the vehicle's motion constraints. These constraints are determined using odometer measurements and posterior estimations from the previous epoch $k-1$, as illustrated in equations (\ref{lat const}) and (\ref{long const}). These equations are derived from the non-holonomic constraints of land vehicles \cite{RISS}.

\begin{equation} \label{lat const}
    \Delta \varphi_{const.} = \frac{\cos r_{k-1}^+ \cos A_{k-1}^+ (v_{Odo_k}+ \epsilon) dt}{R_M + h_{k-1}^+}
\end{equation}

\begin{equation} \label{long const}
    \Delta \lambda_{const.} = \frac{\sin r_{k-1}^+ \cos A_{k-1}^+ (v_{Odo_k}+ \epsilon) dt}{(R_N + h_{k-1}^+) \cos\varphi_{k-1}^+}
\end{equation}

\noindent Where $\epsilon$ denotes the quantization error of the odometer, and $dt$ denotes the sampling time. The SBR measurements are then incorporated in the measurement vector if they satisfy the motion constraint of the vehicle, as shown in (\ref{SBR}).

\begin{equation}\label{SBR}
 \text{SBR}=
\begin{cases}
\text{Include}, &\Delta \varphi<\Delta \varphi_{const.} \land \Delta \lambda<\Delta \lambda_{const.}\\
\text{Discard}, &\text{otherwise.}
\end{cases}
\end{equation}

\noindent Where $\Delta \varphi$ and $\Delta \lambda$ are the geodetic velocities estimated by the SBR measurement and are computed as seen in (\ref{Delta}).

\begin{equation}\label{Delta}
\begin{split}
    \Delta \varphi&=\varphi_{k-1}^+ - \varphi_{k_{SBR}}\\
    \Delta \lambda&=\lambda_{k-1}^+ - \lambda_{k_{SBR}}
\end{split}
\end{equation}

\subsubsection{Observation Model}

The observation model representing the relationship between states and observations is linear, as demonstrated in (\ref{H INS}).

\begin{equation}\label{H INS}
    \boldsymbol{H}= \begin{bmatrix}
        \boldsymbol{I}_{3\times3} & \boldsymbol{0}_{3\times3} & \boldsymbol{0}_{3\times9}\\
        \boldsymbol{0}_{3\times3} & \boldsymbol{I}_{3\times3} & \boldsymbol{0}_{3\times9}     
    \end{bmatrix}
\end{equation}

\subsubsection{Measurement Noise Covariance}
The measurement covariance matrix is shown in (\ref{R}). Entries for positioning that rely on 5G, whether LoS measurements or SBRs are denoted by $\boldsymbol{\sigma}^2_{\varphi_{5G}}$, $\boldsymbol{\sigma}^2_{\lambda_{5G}}$, and $\boldsymbol{\sigma}^2_{h_{5G}}$. 

\begin{equation}\label{R}
\begingroup 
\setlength\arraycolsep{0.5pt}
    \boldsymbol{R}= \text{diag}\left(\begin{bmatrix}
        \boldsymbol{\sigma}^2_{\varphi_{5G}} &\boldsymbol{\sigma}^2_{\lambda_{5G}} &\boldsymbol{\sigma}^2_{h_{5G}} 
        &\sigma^2_{v_{e_{Odo}}} 
        &\sigma^2_{v_{n_{Odo}}} 
        &\sigma^2_{v_{u_{Odo}}}
    \end{bmatrix}\right)
    \endgroup
\end{equation}

\section{Road Tests Setup}
\begin{figure}[h]
	\centering
	\includegraphics[width=\columnwidth]{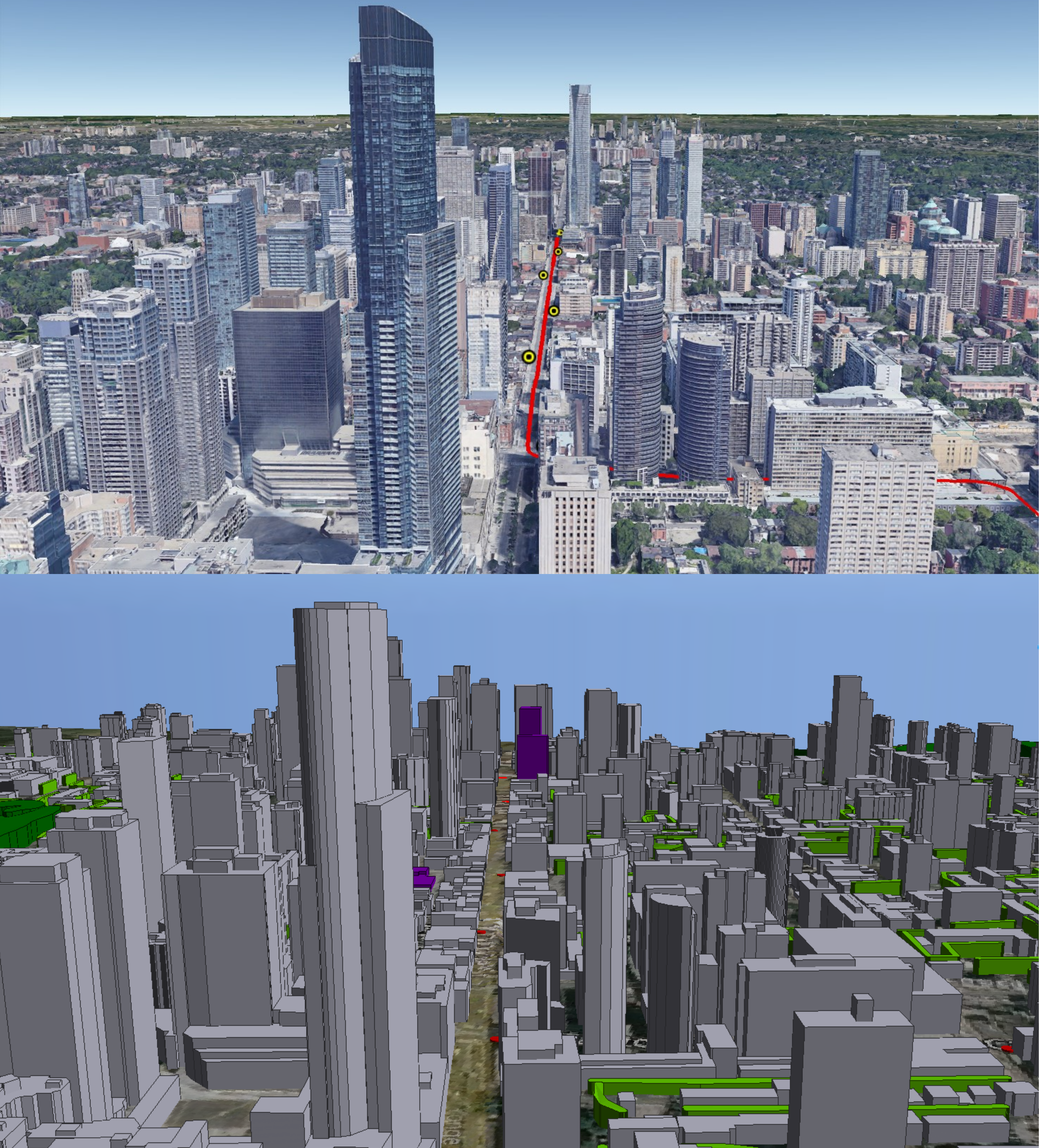}
	\DeclareGraphicsExtensions.
	\caption{Downtown Toronto, ON, Google Earth (Top) vs Siradel simulation tool (Bottom).}
	\label{GoogleEarth vs Siradel}
\end{figure}

A quasi-real 5G simulation configuration offered by Siradel was used for validation. Siradel 5G Channel suite incorporates LiDAR-based maps of the structures, vegetation, and water bodies in downtown regions of cities like Toronto, as shown in Fig. \ref{GoogleEarth vs Siradel}. The simulation tool uses its ray-tracing capabilities and propagation models to calculate necessary positioning measurables like RSS, ToA, AoA, and AoD based on the position of the UE and the virtually connected BSs. A car equipped with NovAtel's high-end positioning solution, which includes a tactical grade KVH 1750 IMU, and a tactical grade GNSS receiver, was driven in Downtown Toronto to simulate a real urban navigation situation. Then, in accordance with the Release 16 guidelines of the 3GPP, BSs were placed approximately $250$ m apart along the driven trajectory. Finally, Siradel was used to create the required 5G measurables using the imported BS positions and NovAtel's reference solution. The mmWave transmissions used by Siradel have a carrier frequency of $28$ GHz and a bandwidth of $400$ MHz. The UE was equipped with an omnidirectional antenna, while the BSs had $8\times1$ ULAs.

\begin{figure}[h]
	\centering
	\includegraphics[width=\columnwidth]{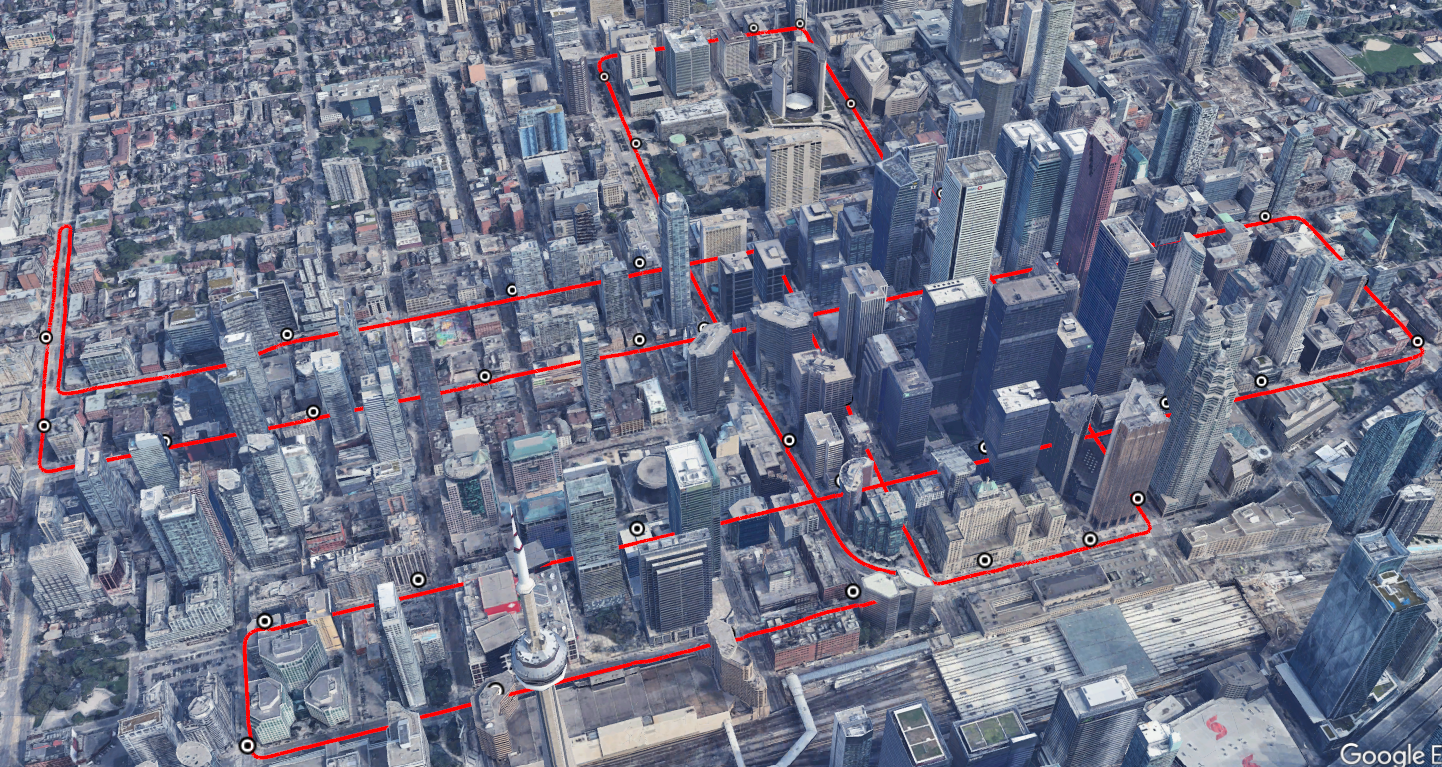}
	\DeclareGraphicsExtensions.
	\caption{Downtown Toronto Trajectory NavINST 1 (Red), and 5G BSs (Yellow circles).}
	\label{Traj12}
\end{figure}

\begin{figure}[h]
	\centering
	\includegraphics[width=\columnwidth]{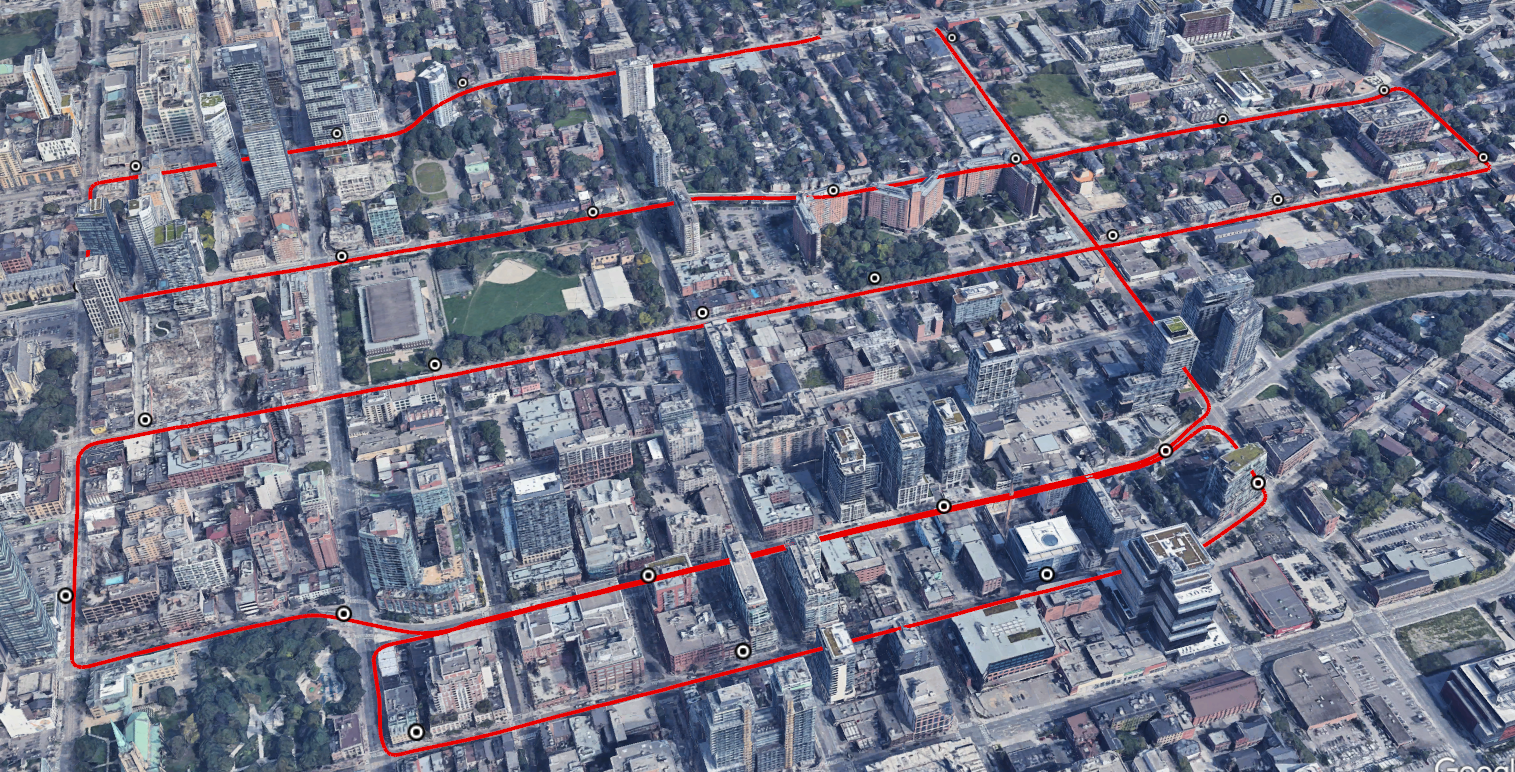}
	\DeclareGraphicsExtensions.
	\caption{Downtown Toronto Trajectory NavINST 2 (Red), and 5G BSs (Yellow circles).}
	\label{Traj43}
\end{figure}

Two test trajectories, namely NavINST 1 and NavINST 2, are used for validation in this work, as seen in Figs. \ref{Traj12} and \ref{Traj43} respectively. The characteristics of each trajectory, along with the equipment used, are summarized in Table \ref{exp}.

\begin{table}[h] \label{exp}
	\caption{Characteristics of trajectories NavINST 1 and NavINST 2}
	\label{equipment}
	\begin{tabularx}{\columnwidth}{@{}l*{3}{C}c@{}}
		\toprule
		&\hspace{-20pt} Sensor     &\hspace{-20pt} NavINST 1 	    &\hspace{-20pt} NavINST 2\\
		\midrule
		&\hspace{-20pt} IMU        &\hspace{-20pt} SCC1300 @ 20 Hz    &\hspace{-20pt} Zed2i IMU @ 50 Hz\\ 
		&\hspace{-20pt} Odometer   & \hspace{-20pt} OBD II @ 1 Hz         &\hspace{-20pt} OBD II @ 3 Hz\\
		&\hspace{-20pt} Distance [km]   &\hspace{-20pt} 9 	           &\hspace{-20pt} 7.5\\ 
		&\hspace{-20pt} Duration [hr]   &\hspace{-20pt} 1.25 	       &\hspace{-20pt} 0.4\\
		\bottomrule
	\end{tabularx}
\end{table}

The trajectories were carried out during rush hour, resulting in numerous instances of sudden car acceleration and stopping dynamics. Furthermore, the trajectories included many turns and challenging maneuvers.

\section{Results and Discussions}
\subsection{Standalone Positioning}
\begin{figure}[h]
	\centering
	\includegraphics[width=\columnwidth]{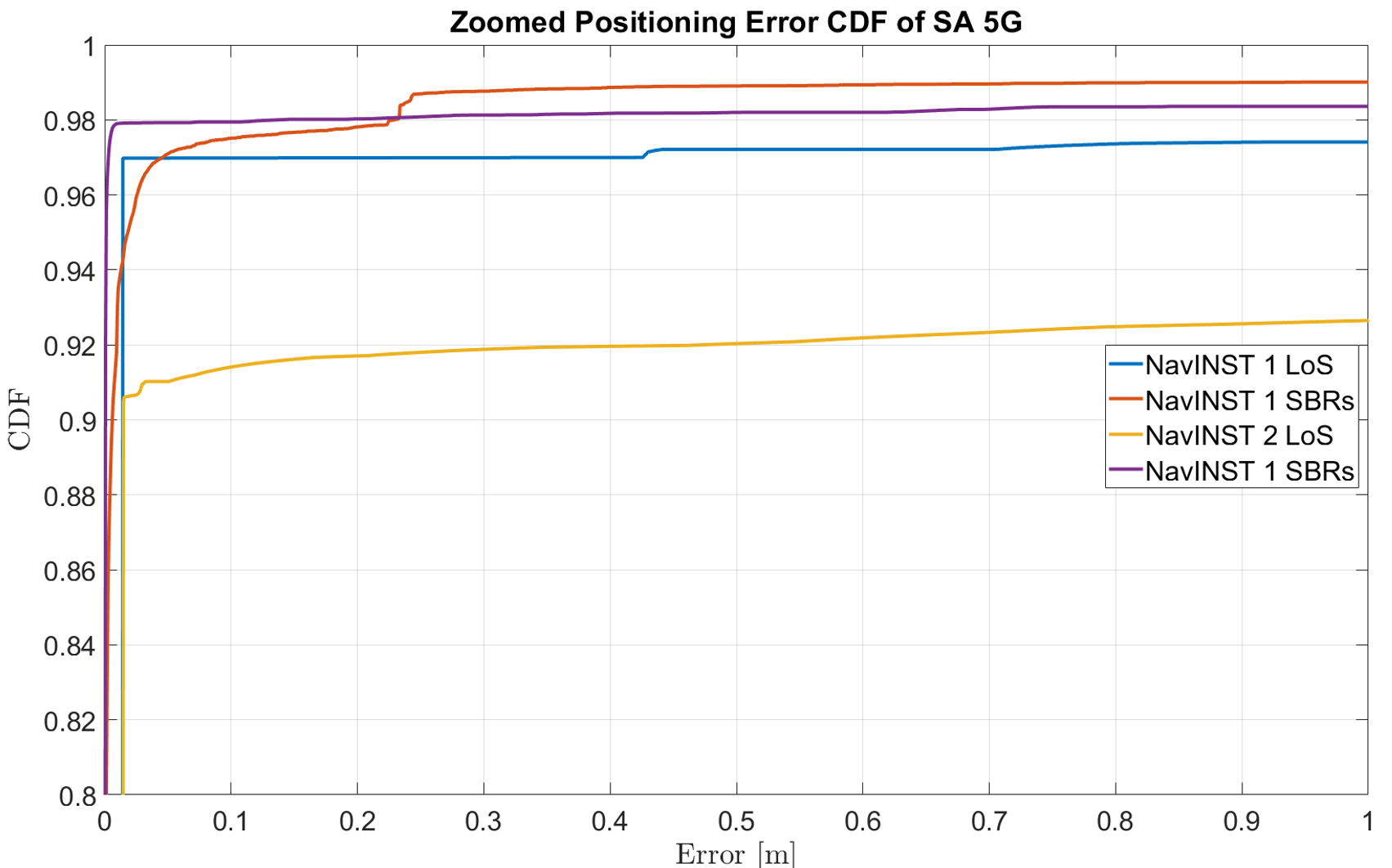}
	\DeclareGraphicsExtensions.
	\caption{CDF of the positioning errors of standalone 5G-LoS (solid) positioning vs. 5G-SBRs (dashed) for trajectories NavINST 1 and NavINST 2.}
    \label{SA 5G CDF}
\end{figure}

This section presents the positioning solution error statistics for the standalone (SA) operation of INS, 5G-LoS, and 5G-SBRs. Tables \ref{table:SAnav1}-\ref{table:SAnav2} summarize the error statistics of trajectories NavINST 1 and 2, respectively. Fig. \ref{SA 5G CDF} shows the error cumulative distribution function (CDF) of all 5G SA positioning solutions. In Table \ref{table:SAnav1}, it can be seen that 5G LoS- and SBRs-based positioning have close error statistics, with SBRs providing slightly better results when they are available. However, their RMS and max errors are drastically higher as they cause severe positioning errors when the available SBRs are insufficient (i.e. less than two). The dissimilarity in error statistics is evident from Table \ref{table:SAnav2}, where the trajectory exhibits a reduced likelihood of LoS communication with the BS. This finding indicates that the probability of obtaining a sufficient number of SBRs in urban settings is higher than the probability of LoS communication.
\begin{table}[h] 

	\caption{2D Standalone Positioning Error Statistics Summary for Trajectory NavINST 1}
         \label{table:SAnav1}
	\begin{tabularx}{\columnwidth}{@{}l*{5}{C}c@{}}
		\toprule
		&Statistics &SA INS	      &SA 5G-LoS  &SA 5G-SBRs\\
		\midrule
		&RMS        & 450 km      &6.3 m     &40 m\\ 
		&Max        & 750 km      &107 m     &4734 m\\
		&Sub-$2$ m      & 0.6\% 	  &97.4\%     &99\%\\ 
		&Sub-$1$ m      & 0.4\% 	  &97.4\%     &99\%\\
		&Sub-$30$ cm   & 0.2\%       &97\%       &98.8\%\\
		\bottomrule
	\end{tabularx}
\end{table}

\begin{table}[h] 
	\caption{2D Standalone Positioning Error Statistics Summary for Trajectory NavINST 2}
         \label{table:SAnav2}
	\begin{tabularx}{\columnwidth}{@{}l*{5}{C}c@{}}
		\toprule
		&Statistics &SA INS	      &SA 5G-LoS  &SA 5G-SBRs\\
		\midrule
		&RMS        & 40 km      &4 m      &27 m\\ 
		&Max        & 65 km    &64 m     &3897 m\\
		&Sub-$2$ m      & 0.8\% 	  &94\%      &98.4\%\\ 
		&Sub-$1$ m      & 0.06\% 	  &93\%      &98.4\%\\
		&Sub-$30$ cm   & 0.02\%      &92\%      &98.1\%\\
		\bottomrule
	\end{tabularx}
\end{table}

\begin{figure}[h]

	\centering
	\includegraphics[width=\columnwidth]{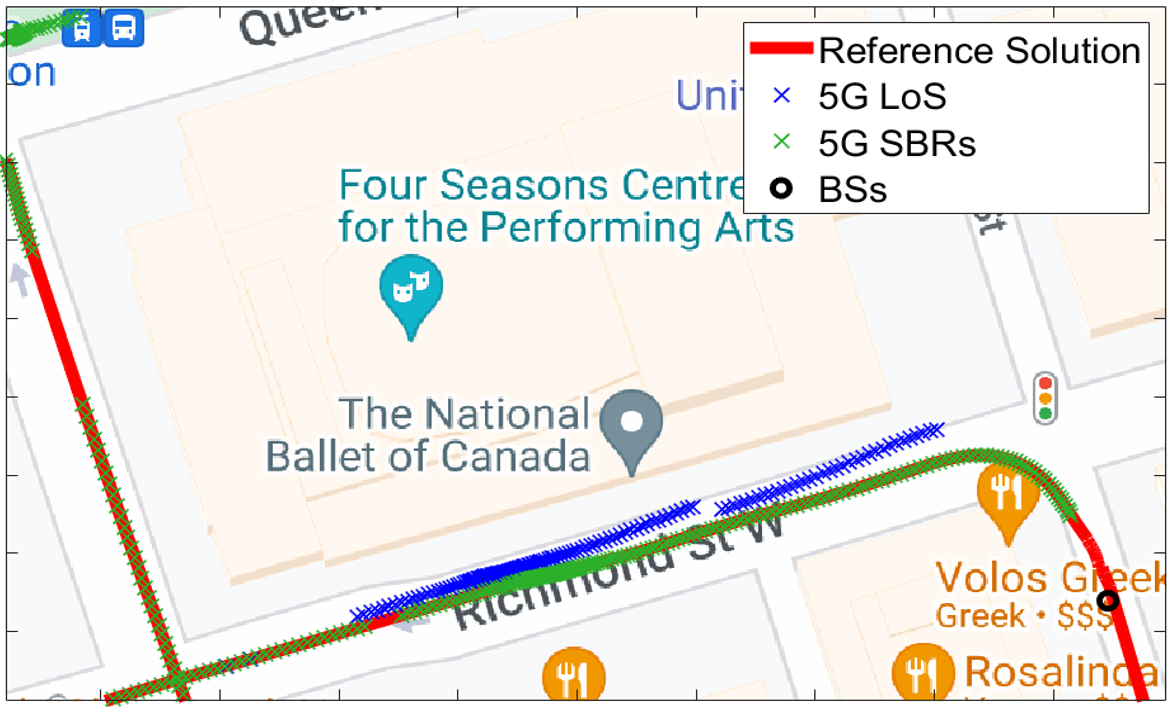}
	\DeclareGraphicsExtensions.
	\caption{Close-up scenario that showcases the capability of multipath positioning accuracy during LoS outage.}
         \label{CU-1}
\end{figure}

\begin{figure}[h] 
	\centering
	\includegraphics[width=\columnwidth]{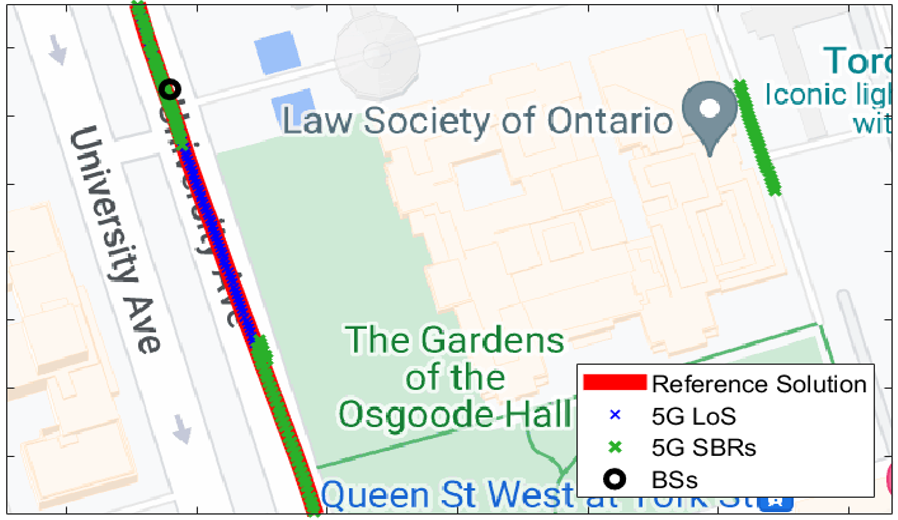}
	\DeclareGraphicsExtensions.
	\caption{Close-up scenario that showcases the positioning solution of utilizing LoS measurements during SBRs outage.}
         \label{CU-2}
\end{figure}

\begin{figure}[h] 
	\centering
	\includegraphics[width=\columnwidth]{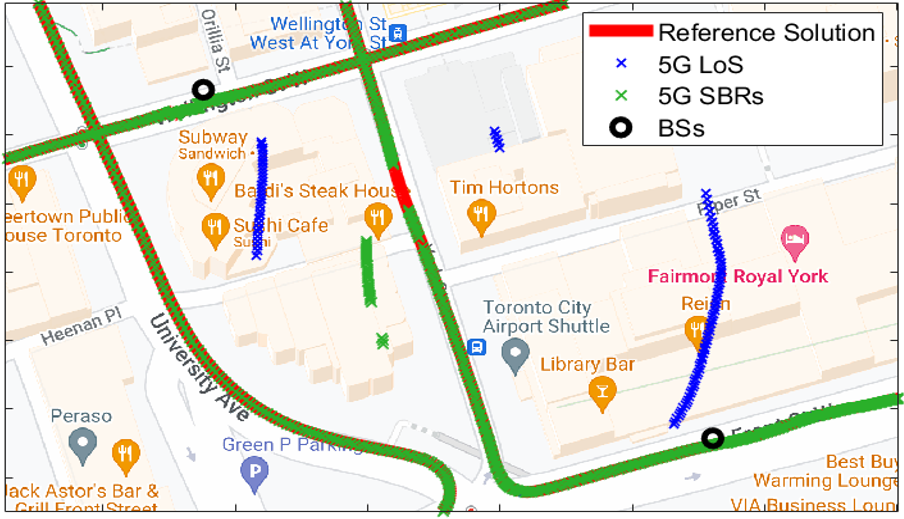}
	\DeclareGraphicsExtensions.
	\caption{Close-up scenario that shows an instance where both LoS and SBRs are not available.}
        \label{CU-3}
\end{figure}

Close-ups of the LoS and SBRs-based positioning solution are shown in Figs. \ref{CU-1}-\ref{CU-2}. It can be seen that LoS and multipath signals complement each other when either of them is not available. Such dynamic necessitates the integration between them to achieve a higher percentage of sub-$30$ cm level of accuracy, as well as contained RMS and max errors. However, in some instances, as shown in Fig. \ref{CU-3}, both LoS and two SBRs are unavailable, resulting in a total 5G outage. As a result, the integration with OBMS to bridge these gaps become necessary. 

\subsection{5G LoS Positioning Integrated with OBMS} \label{sec2}
\begin{figure}[h] 
	\centering
	\includegraphics[width=\columnwidth]{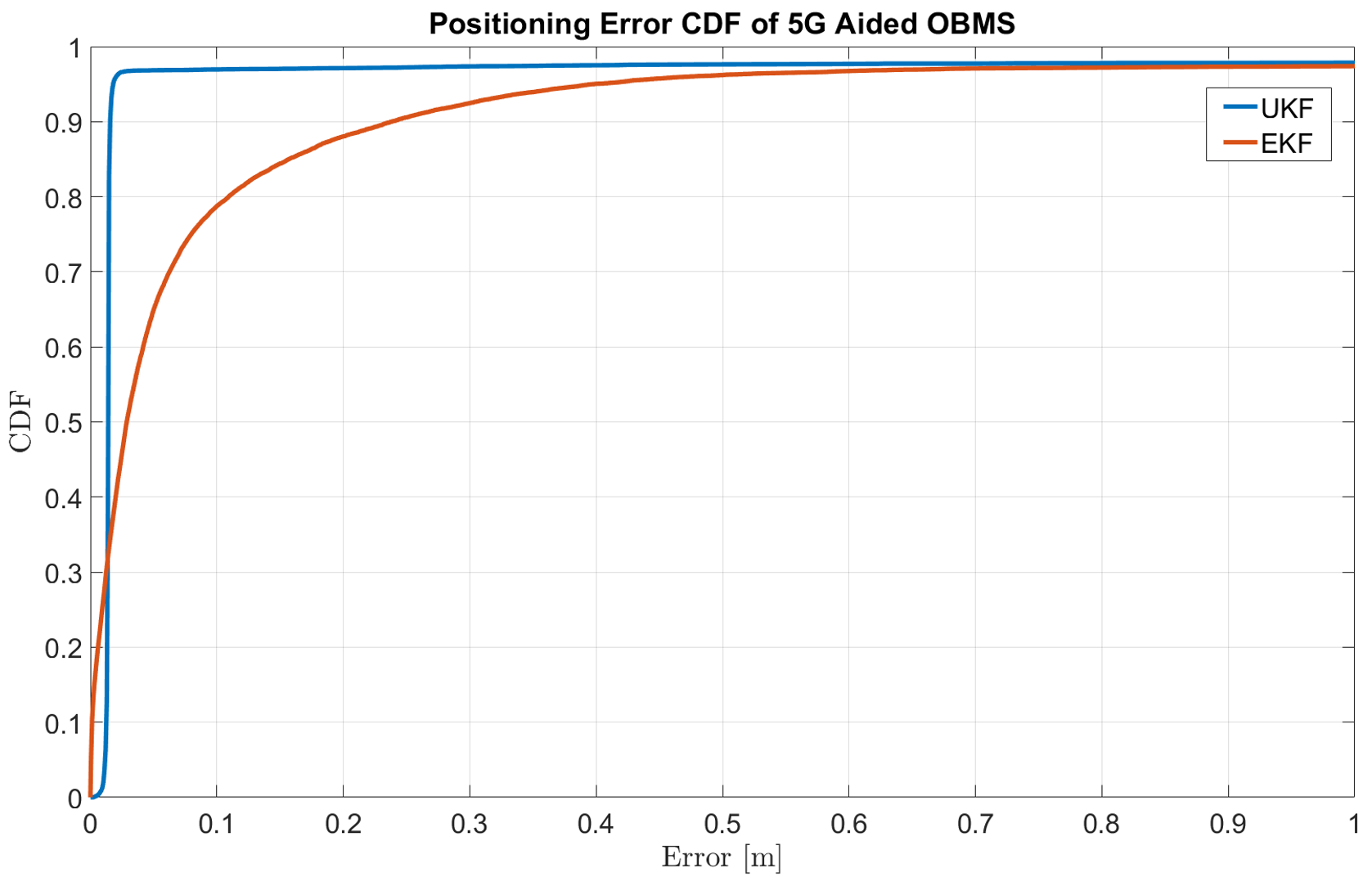}
	\DeclareGraphicsExtensions.
	\caption{CDF of the positioning errors of 5G aided OBMS positioning using EKF vs. UKF for trajectory NavINST 1.}
        \label{cdf sec2_1}
\end{figure}

\begin{figure}[h] 
	\centering
	\includegraphics[width=\columnwidth]{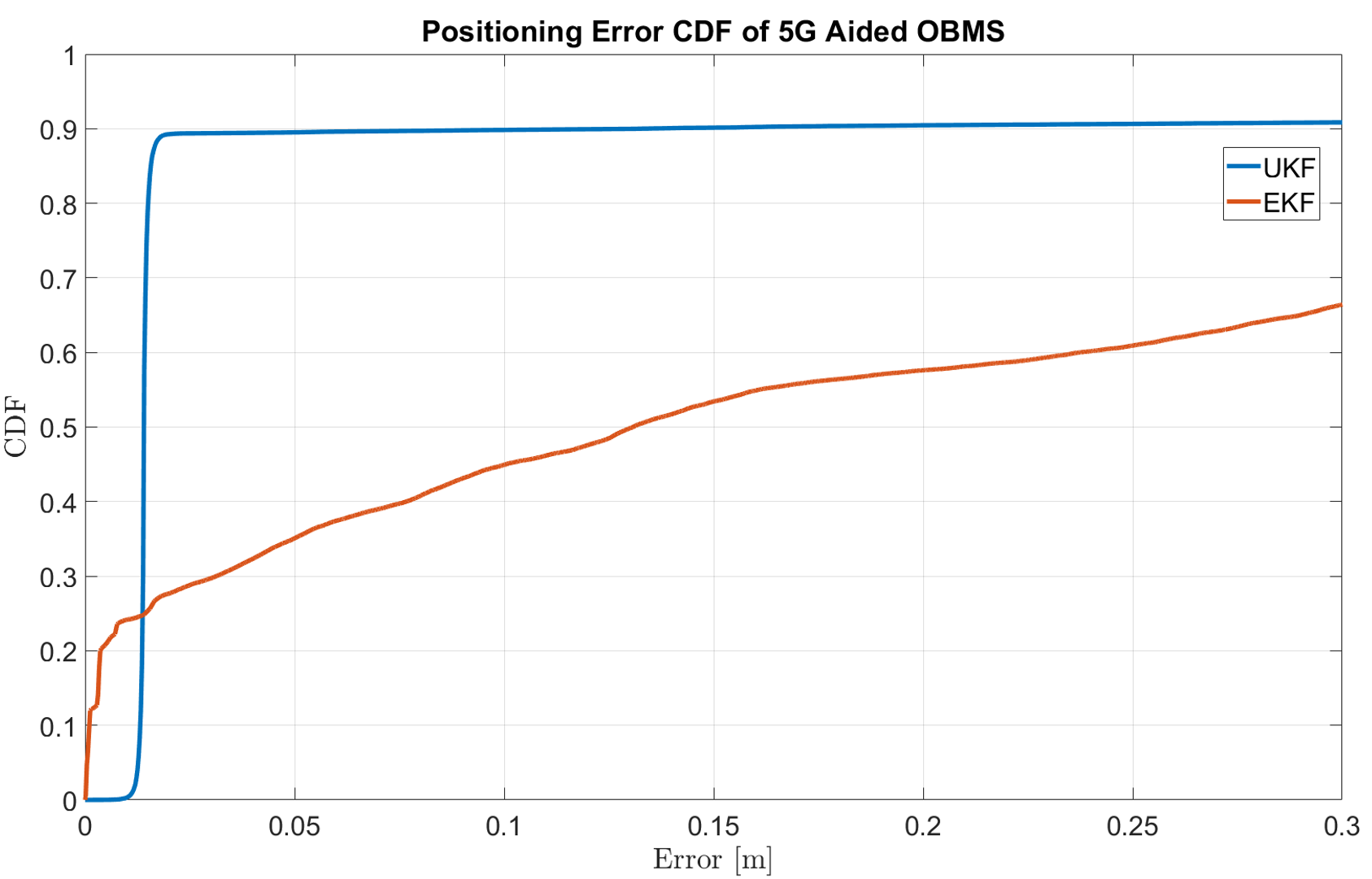}
	\DeclareGraphicsExtensions.
	\caption{CDF of the positioning errors of 5G aided OBMS positioning using EKF vs. UKF for trajectory NavINST 2.}
        \label{cdf sec2_2}
\end{figure}

This section introduces the initial stage of integrating 5G LoS measurements with OBMS. Our primary objective is to assess the effectiveness of the proposed UKF as a fusion engine in contrast to the commonly employed EKF. Table \ref{table:sec2} presents a summary of the error statistics after integration using both UKF and EKF for trajectories NavINST 1 and NavINST 2. Overall, it is evident that UKF is delivering superior outcomes when compared to EKF. This can be attributed to the linearization errors that occur in EKF due to linearizing the state transition and observation models. This error leads to a less accurate prediction of the system covariance matrix $\boldsymbol{P}$ and computation of the Kalman gain $\boldsymbol{K}$, both of which contribute to poor state estimates. However, a significant difference between the two fusion systems solutions is more apparent in the NavINST 2 trajectory than in NavINST 1, as seen in Figs. \ref{cdf sec2_1} and \ref{cdf sec2_2}. One possible interpretation of these results is that, in general, NavINST 2 trajectory exhibits a higher frequency and longer duration of 5G outages, as indicated in Tables \ref{table:SAnav1} and \ref{table:SAnav2}. This may be compounded by the use of a poor IMU in NavINST 2 to bridge these gaps.
\begin{table}[h]
	\caption{2D Positioning Error Statistics of 5G Aided OBMS using EKF vs. UKF.}
	\label{table:sec2}
        \centering
	\begin{tabularx}{\columnwidth}{@{}l*{5}{c}c@{}}
		\toprule
		&Error      &\multicolumn{2}{c}{NavINST 1}     &\multicolumn{2}{c}{\hspace{12pt}NavINST 2}\\    
        &\hspace{12pt}Type\hspace{12pt}         &\hspace{12pt}EKF\hspace{12pt}         &\hspace{12pt}UKF\hspace{12pt}           &\hspace{12pt}EKF\hspace{12pt}     &UKF\\
		\midrule
		&RMS            & $1.8$ m     & $0.7$ m        &$5.6$ m   &$0.6$ m\\ 
		&Max            & $23$ m      & $8.3$ m        &$51$ m  &$8.7$ m\\ 
		&Sub-$2 $ m      & $98\%$     & $98\%$        &$91\%$  &$99\%$\\ 
		&Sub-$1 $ m      & $97.3\%$   & $98\%$        &$86\%$  &$98\%$\\ 
		&Sub-$30$ cm    & $92.4\%$   & $97.3\%$      &$66\%$  &$96.6\%$ \\
		\bottomrule
	\end{tabularx}
\end{table}

\begin{figure}[h] 
	\centering
	\includegraphics[width=\columnwidth]{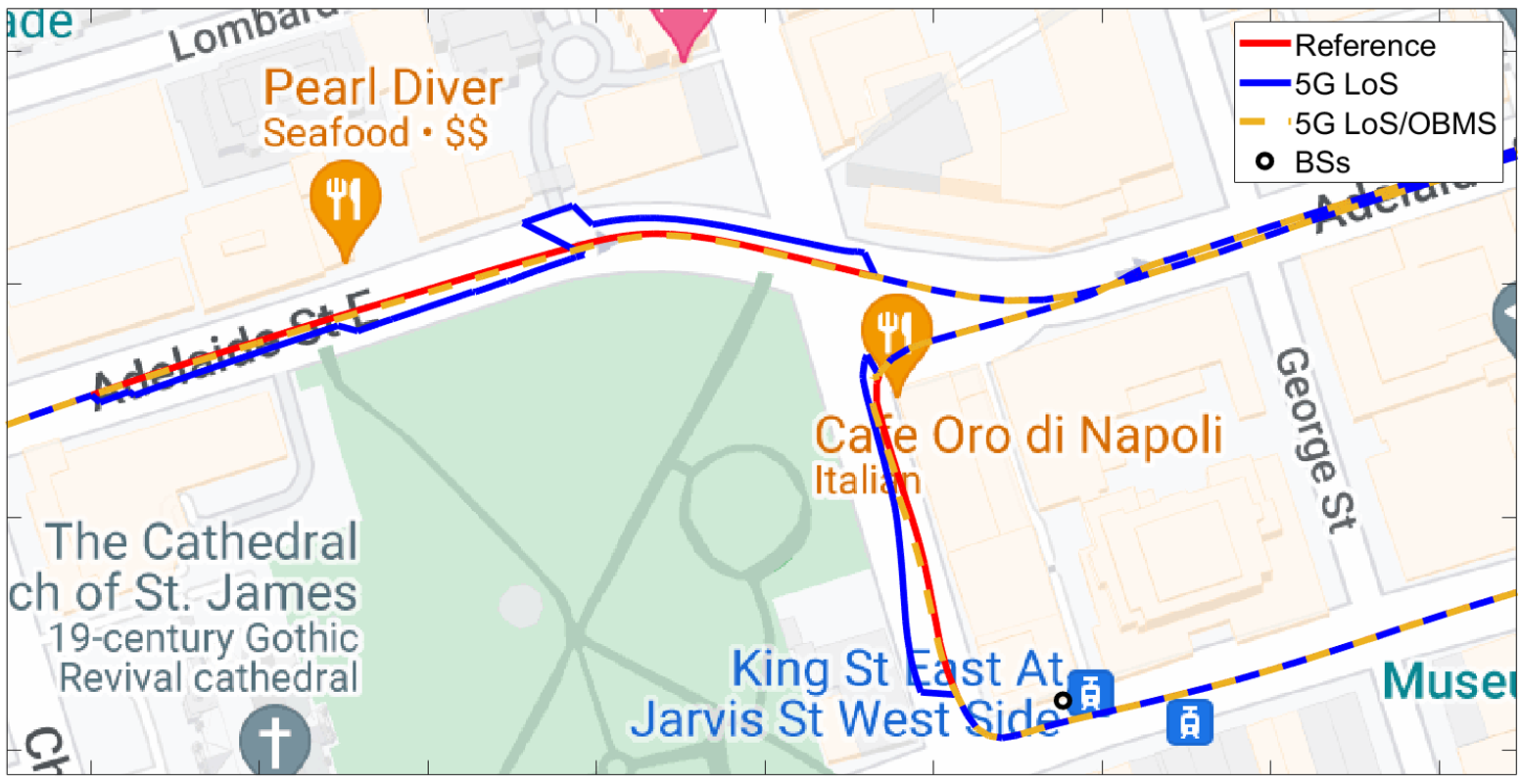}
	\DeclareGraphicsExtensions.
	\caption{A close-up of 5G LoS aided OBMS positioning solution using UKF vs. SA 5G-LoS positioning solution.}
        \label{sec2_map}
\end{figure}

A close-up of the positioning solution of the proposed integration using UKF compared to that of SA 5G LoS measurements is shown in Fig. \ref{sec2_map}.

Another close-up is shown in Fig. \ref{sec2_map} where the 5G LoS outage as previously observed in Fig. \ref{CU-3} has been successfully bridged with the aid of OBMS.

\begin{figure}[h] 
	\centering
	\includegraphics[width=\columnwidth]{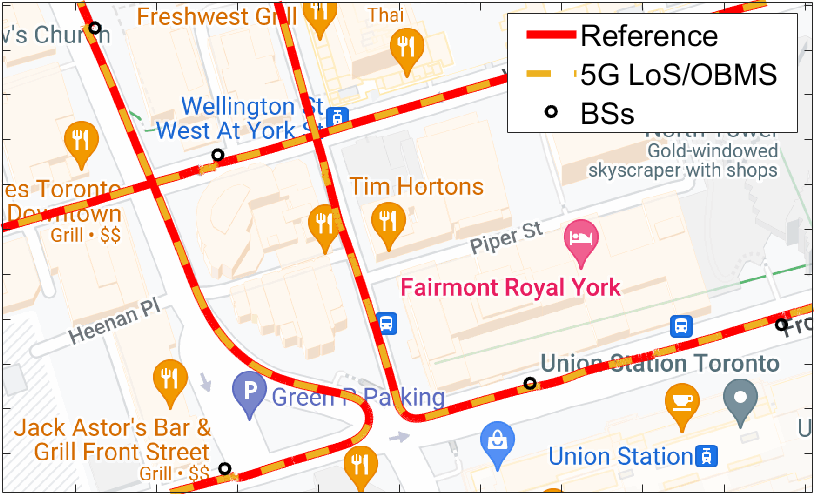}
	\DeclareGraphicsExtensions.
	\caption{A close-up of 5G LoS aided OBMS positioning solution using UKF vs. SA 5G-LoS positioning solution.}
        \label{sec2_map2}
\end{figure}

\subsection{Integration with SBR-based Positioning}
\begin{figure}[h] 
	\centering
	\includegraphics[width=\columnwidth]{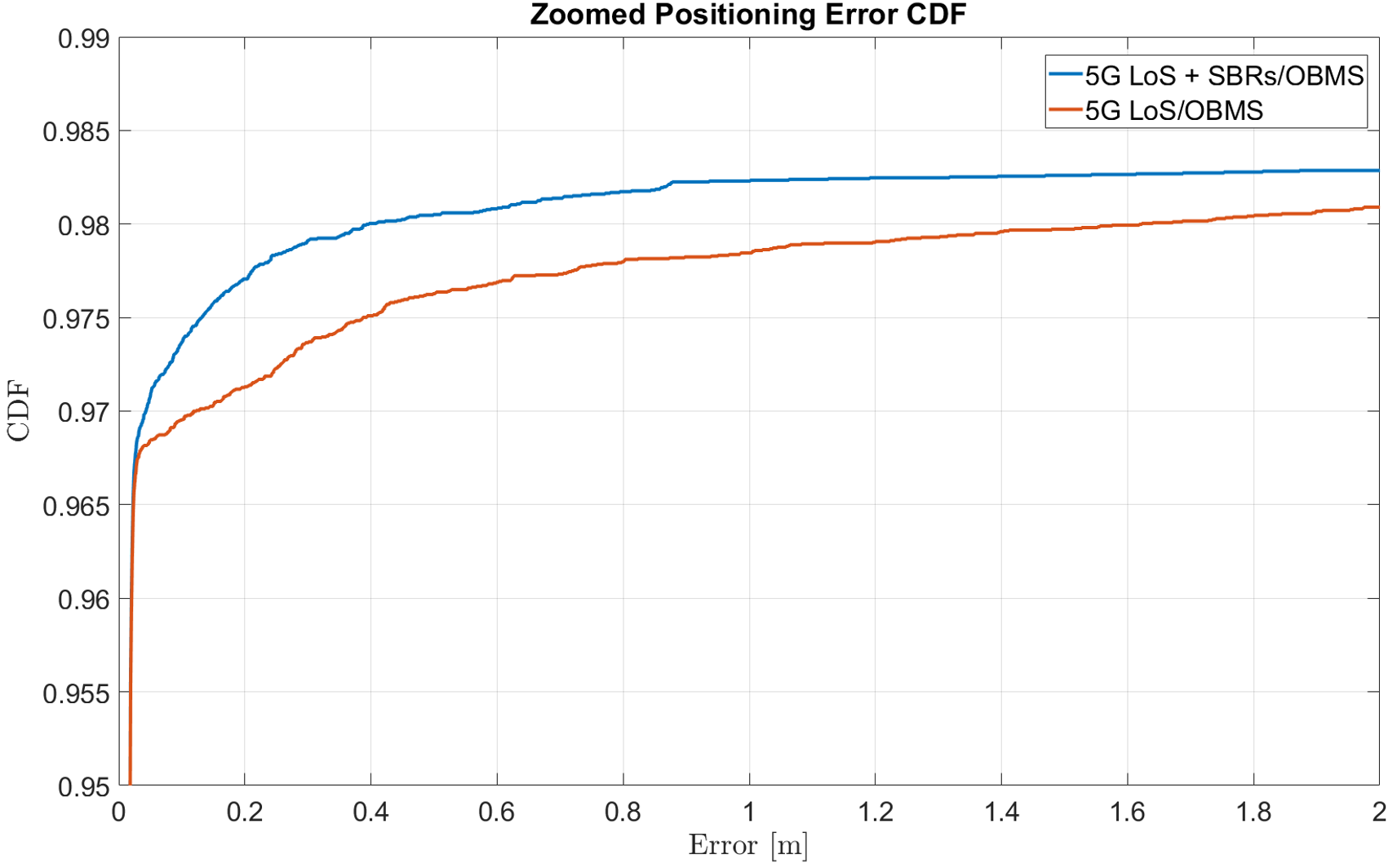}
	\DeclareGraphicsExtensions.
	\caption{CDF of the positioning errors of 5G aided OBMS positioning with and without SBRs for trajectory NavINST 1.}
        \label{cdf sec3_1}
\end{figure}

\begin{figure}[h] 
	\centering
	\includegraphics[width=\columnwidth]{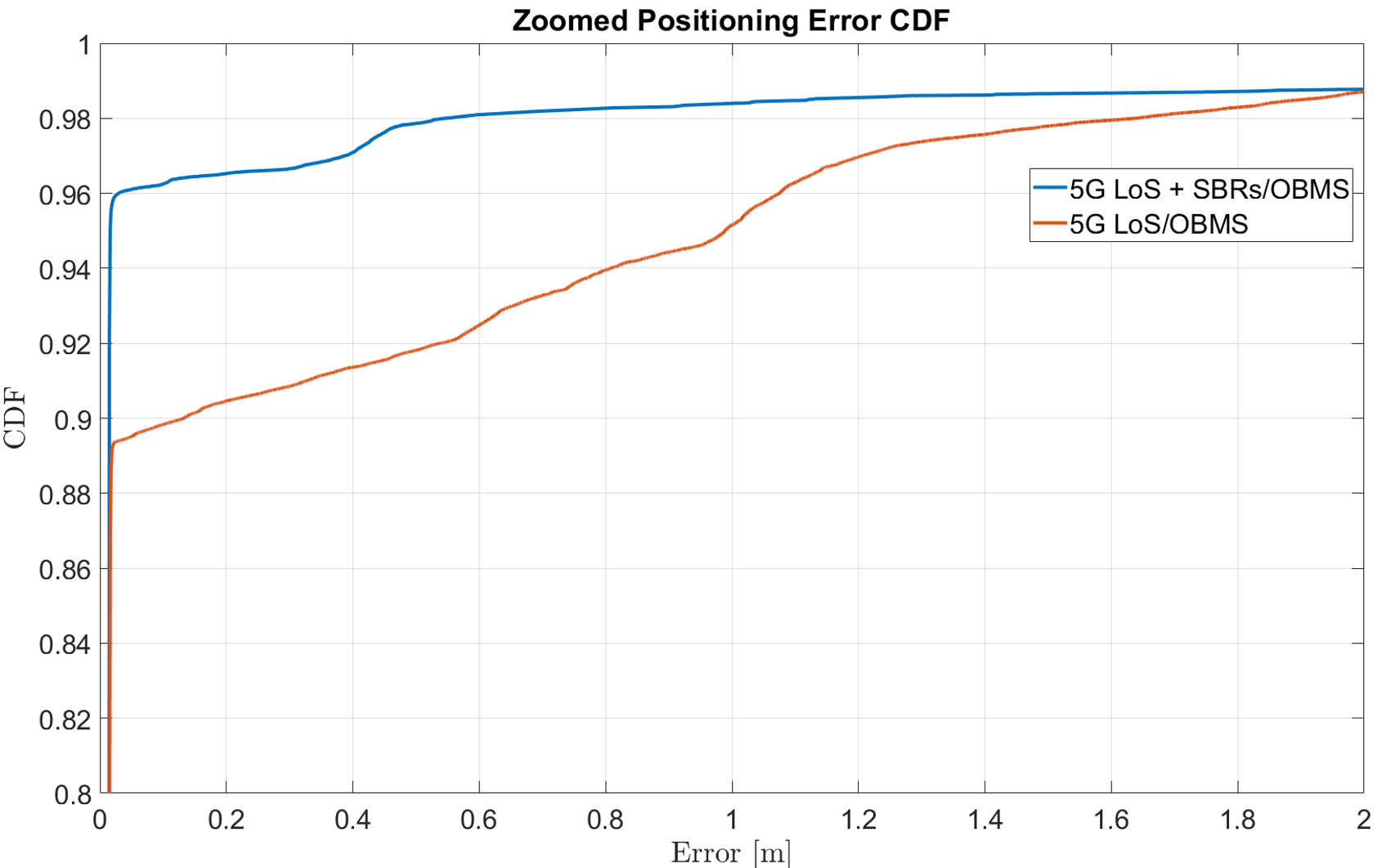}
	\DeclareGraphicsExtensions.
	\caption{CDF of the positioning errors of 5G aided OBMS positioning with and without SBRs for trajectory NavINST 2.}
        \label{cdf sec3_2}
\end{figure}

In this section, we expand our integration approach to incorporate SBRs using UKF, building upon earlier findings. The results are summarized in Table \ref{table:sec3}. Furthermore, the positioning error CDF is shown in Figs. \ref{cdf sec3_1}-\ref{cdf sec3_2}. Once again, it is evident that the disparity in results is more pronounced in the NavINST 2 trajectory than in NavINST 1, primarily due to the more frequent occurrence of outages in NavINST 2. Upon examining the results of NavINST 2, it is apparent that integrating multipath signals can maintain a level of accuracy below $30$ cm for $97\%$ of the time, compared to only $91\%$ without utilizing multipath. As a benchmark, reliable operation of autonomous vehicles requires a decimeter level positioning accuracy of $<30$ cm for at least $2\sigma$, $(>95\%)$ of the time \cite{AVRequirements}.
\begin{table}[h]
	\caption{2D Positioning Error Statistics of 5G Aided OBMS with and without integration with SBRs.}
	\label{table:sec3}
        \centering
	\begin{tabularx}{\columnwidth}{@{}l*{5}{c}c@{}}
		\toprule
		&Error      &\multicolumn{2}{c}{NavINST 1}     &\multicolumn{2}{c}{\hspace{3pt}NavINST 2}\\    
        &\hspace{3pt}Type\hspace{3pt}         &\hspace{3pt}W/o SBRs\hspace{3pt}         &\hspace{3pt}W/ SBRs\hspace{3pt}           &\hspace{3pt}W/o SBRs\hspace{3pt}     &W/ SBRs\\
		\midrule
		&RMS            & $0.7$ m      & $0.2$ m        &$0.5$ m    &$0.2$ m\\ 
		&Max            & $8.3$ m      & $4.5$ m        &$11.7$ m    &$3.1$ m\\ 
		&Sub-$2 $ m      & $98\%$      & $98.2\%$        &$98.7\%$    &$99\%$\\ 
		&Sub-$1 $ m      & $97.8\%$      & $98.2\%$        &$95\%$    &$98.5\%$\\ 
		&Sub-$30$ cm  & $97.3\%$    & $98\%$      &$90.8\%$  &$96.3\%$ \\
		\bottomrule
	\end{tabularx}
\end{table}

\begin{figure}[h] 
	\centering
	\includegraphics[width=\columnwidth]{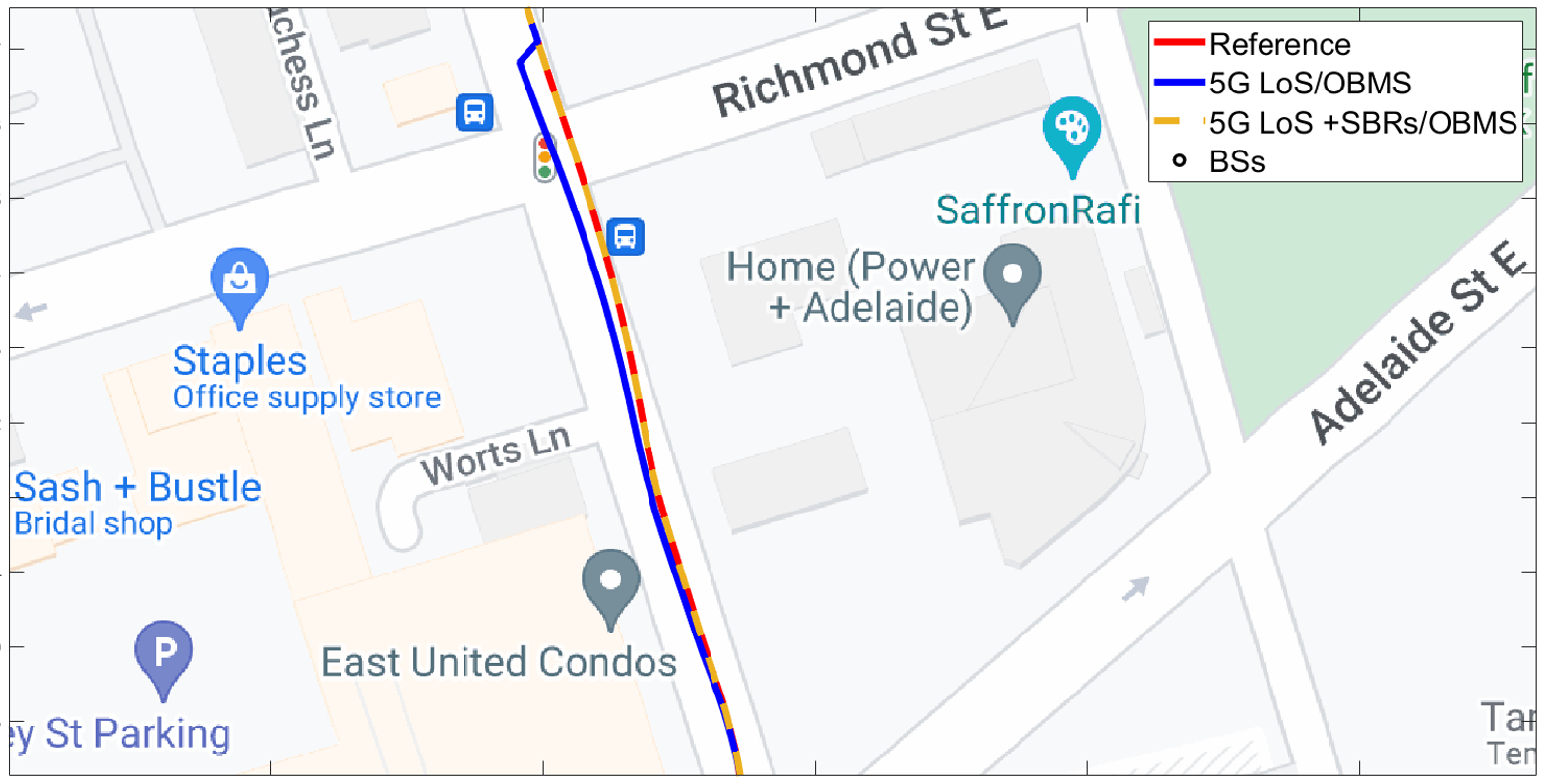}
	\DeclareGraphicsExtensions.
	\caption{A close-up of 5G LoS aided OBMS positioning solution with and without SBRs.}
        \label{sec3_map}
\end{figure}

\begin{figure}[h] 
	\centering
	\includegraphics[width=\columnwidth]{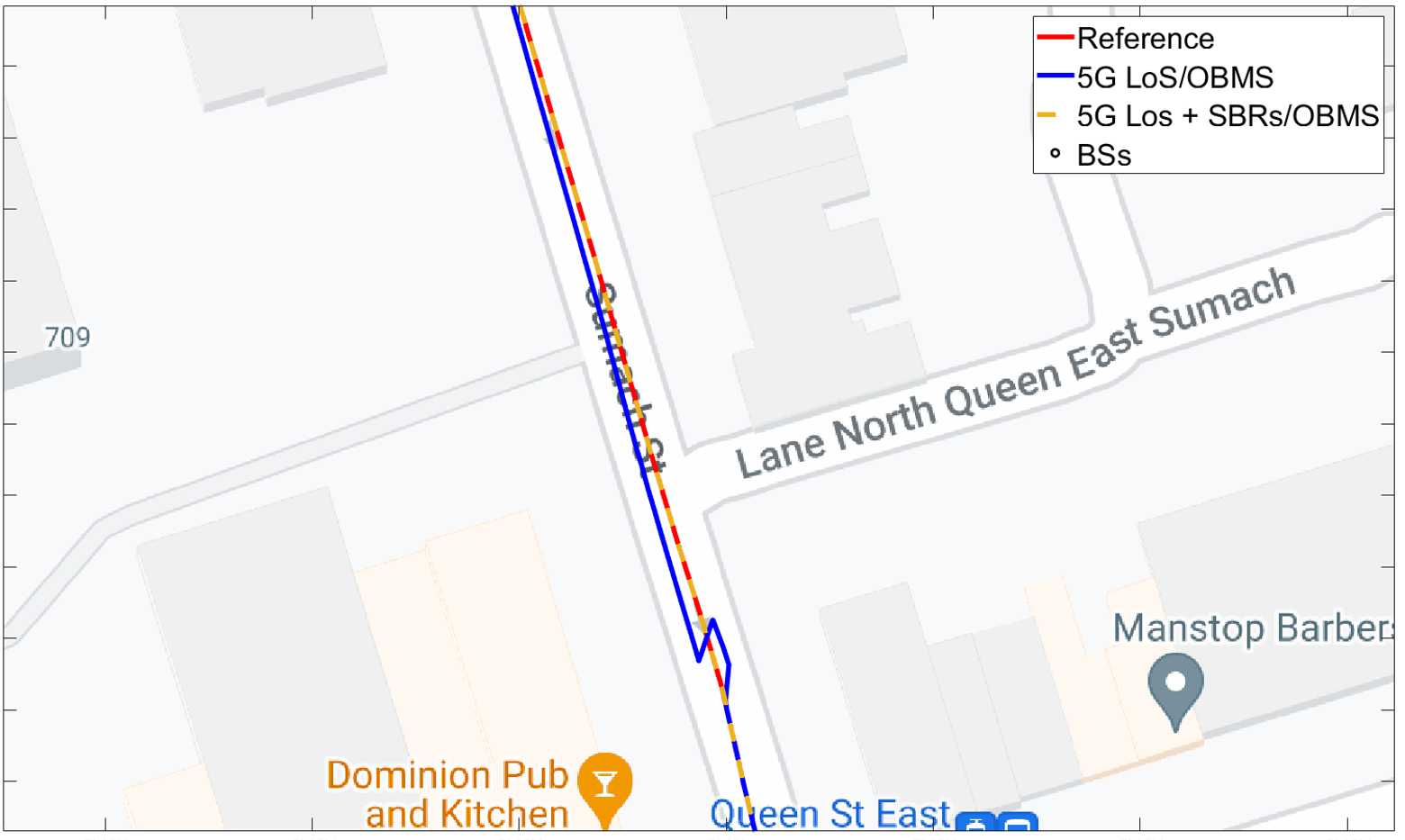}
	\DeclareGraphicsExtensions.
	\caption{A close-up of 5G LoS aided OBMS positioning solution with and without SBRs (trajectory NavINST 2).}
        \label{sec3_map2}
\end{figure}

\begin{figure}[h] 
	\centering
	\includegraphics[width=\columnwidth]{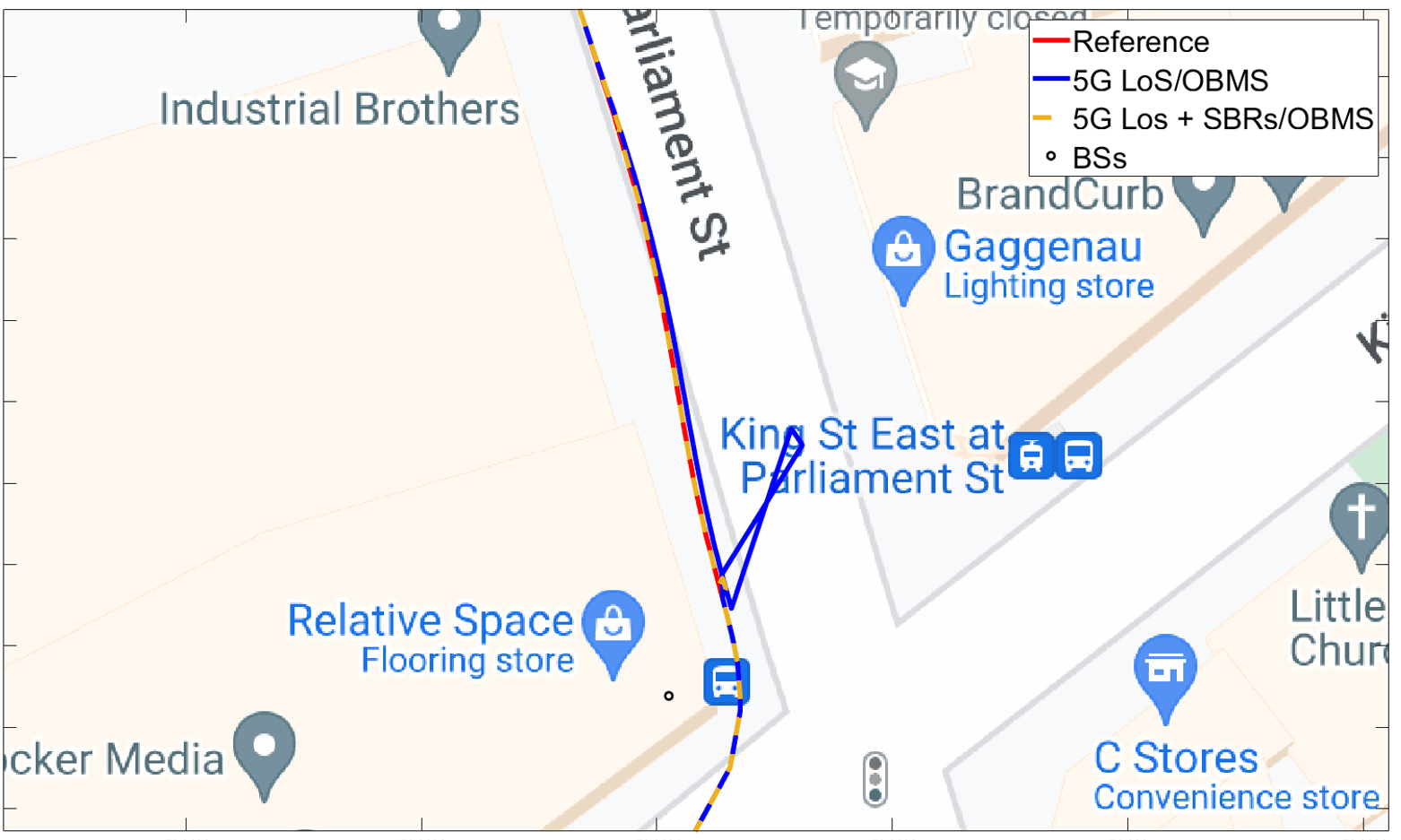}
	\DeclareGraphicsExtensions.
	\caption{A close-up of 5G LoS aided OBMS positioning solution with and without SBRs (trajectory NavINST 2).}
        \label{sec3_map3}
\end{figure}

Figs. \ref{sec3_map}-\ref{sec3_map3} show close-up comparisons between the positioning solution of the proposed integration using UKF with and without SBRs. The results indicate that prolonged LoS outages should be bridged since the IMU positioning solution is prone to drift. Multipath signals are more likely to be present than LoS communication and can serve as a bridge to fill these gaps.

\section{Conclusion}
In conclusion, this paper presents an improved positioning solution for AVs that incorporates 5G mmWave LoS and multipath signals as well as integration with OBMS. The work employs a UKF fusion engine as an alternative to the commonly used EKF. To evaluate the health of the 5G measurements, two techniques were used. The first was based on the communication link between the BS and the UE, while the second relied on the motion restrictions of the vehicle. To validate the proposed methods, two trajectories with real-vehicle dynamics and different low-end IMU units were utilized. A novel quasi-real 5G simulator with ray-tracing capabilities was used to obtain 5G measurements. In the course of our analysis, it was observed that SBRs are more easily accessible compared to LoS links. Moreover, it was found that UKF outperforms EKF, particularly during extended periods of 5G outages.  Finally, we demonstrated the integration capabilities with multipath measurements. Our findings indicate that exploiting available multipath signals is necessary to achieve decimeter-level accuracy. With the proposed positioning solution, the system achieved a sub-$30$ cm level of accuracy for about $97\%$ of the time, compared to only $91\%$ of the time without incorporating multipath signals.



\bibliographystyle{IEEEtran}
\bibliography{References}

\begin{IEEEbiography}[{\includegraphics[width=1in,height=1.25in,clip]{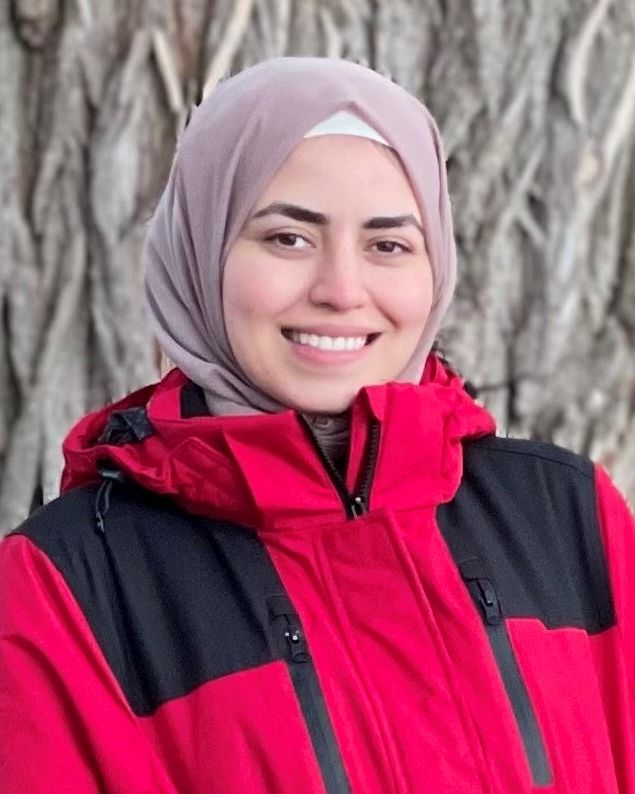}}]{Qamar Bader} (Graduate Student Member, IEEE) received her B.Sc. degree in electrical engineering from Qatar University, Doha, Qatar, in 2016, followed by an MSc degree in electrical engineering at Queen's University, Canada. She is currently pursuing her Ph.D. studies at Queen's University. She is also a member of the Navigation and Instrumentation Research Lab at the Royal Military College of Canada, RMCC. Her current research interests include 5G positioning and navigation, sensor fusion, deep learning, computer vision, and environment mapping.
\end{IEEEbiography}

\begin{IEEEbiography}[{\includegraphics[trim=0 700 0 300, width=1in,height=1.25in,clip]{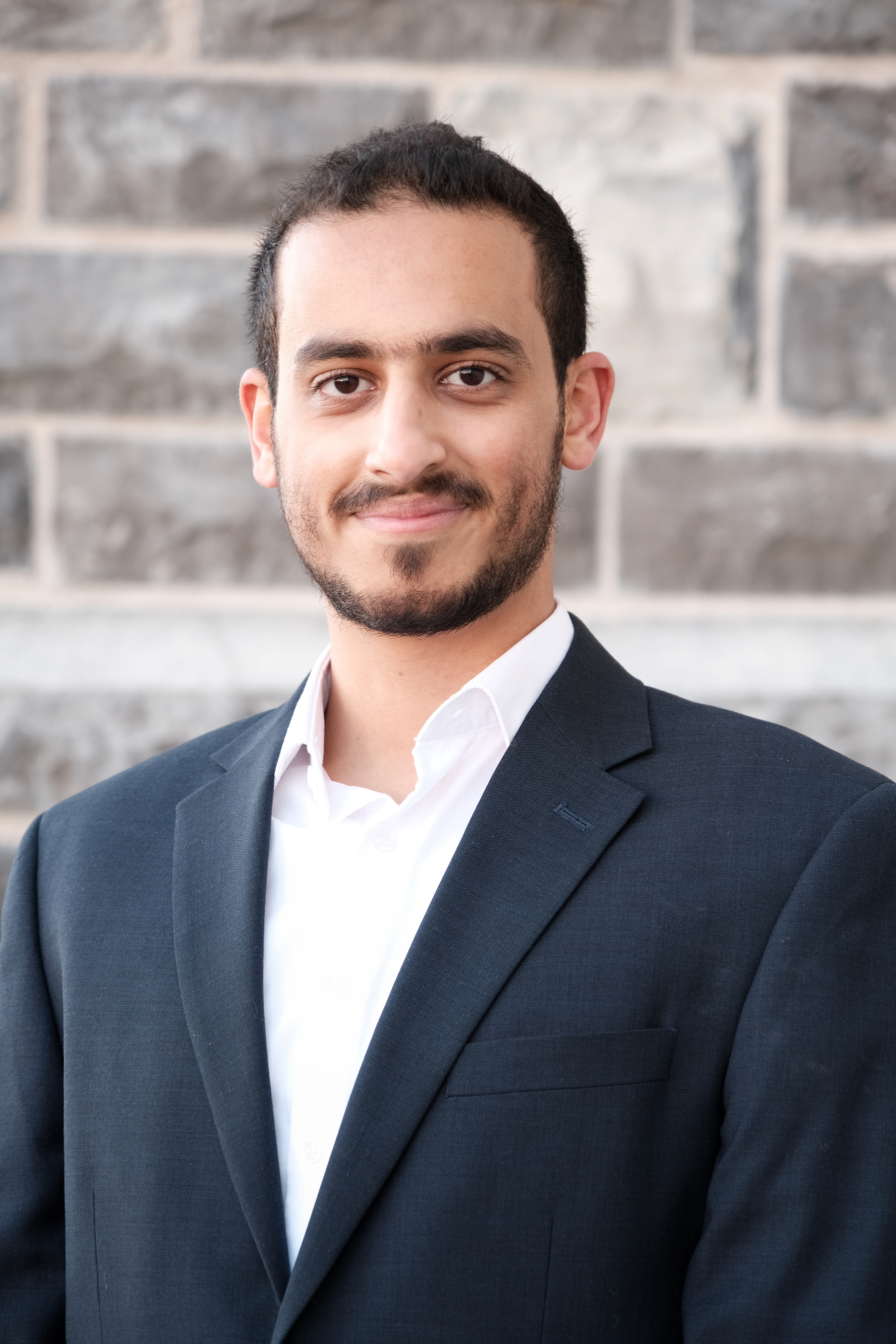}}]{Sharief Saleh} (Graduate Student Member, IEEE) received the B.Sc. and M.Sc. degrees in electrical engineering from Qatar University, Doha, Qatar, in 2016 and 2018 respectively. He completed his Ph.D. degree in electrical engineering at Queen's University, Canada. He was awarded a Graduate Assistant position at Qatar University during his master's studies and was then appointed as a Research Assistant at Qatar University, Doha, Qatar. He is currently a member of the Navigation and Instrumentation Research Lab, RMCC. His current research interests include 5G positioning and navigation, sensor fusion, sensors and instrumentation, signal processing, reinforcement learning, and AI.
\end{IEEEbiography}

\begin{IEEEbiography}[{\includegraphics[width=1in,height=1.25in,clip]{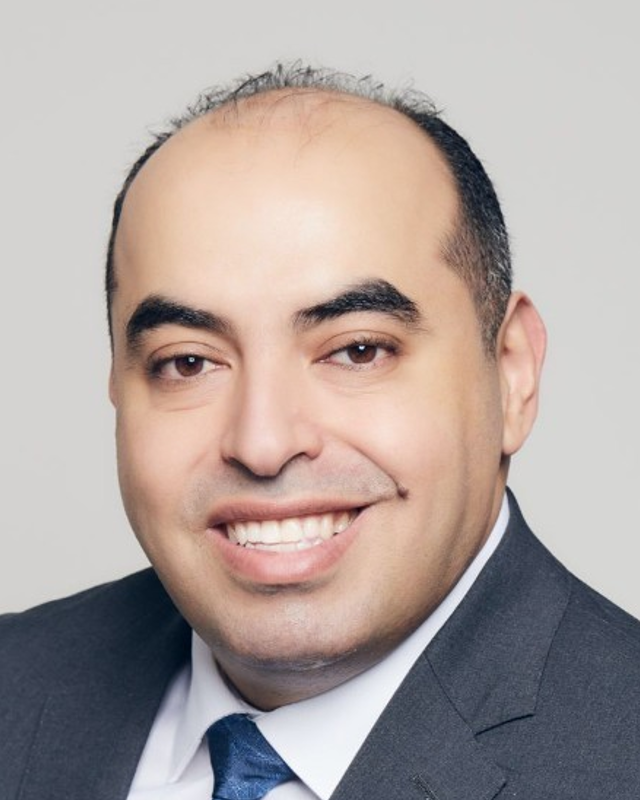}}]{Mohamed Elhabiby} was the Treasurer of the Geodesy Section at the Canadian Geophysical Union from 2008 to 2014. He is currently an Associate Professor with the Faculty of Engineering, Ain Shams University, Cairo, Egypt. He is also the Executive Vice President and Co-Founder of Micro Engineering Tech Inc., Calgary, AB, a high-tech international company, specialized in high-precision engineering and instrumentation, mobile mapping, laser scanning, deformation monitoring, and GPS/INS integrations. He is a Leader of an Archaeological Mission in the Area of the Great Pyramids, Cairo. He received the Astech Awards. He is named by Avenue Magazine as one of the Top 40 under 40. He is the Chair of WG 4.1.4: Imaging Techniques, Sub-Commission 4.1: Alternatives and Backups to GNSS. He chaired the Geocomputations and Cyber Infrastructure Oral Session at the Canadian Geophysical Union annual meeting from 2008 to 2012.
\end{IEEEbiography}
\vfill

\begin{IEEEbiography}[{\includegraphics[trim=0 0 0 0, width=1in,height=1.25in,clip]{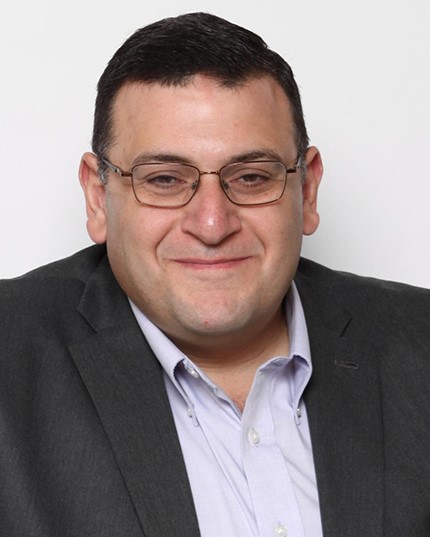}}]{Aboelmagd Noureldin} (Senior Member, IEEE) received the B.Sc. degree in electrical engineering and the M.Sc. degree in engineering physics from Cairo University, Egypt, in 1993 and 1997, respectively, and the Ph.D. degree in electrical and computer engineering from the University of Calgary, AB, Canada, in 2002. He is a Professor at the Department of Electrical and Computer Engineering, Royal Military College of Canada (RMCC), with Cross Appointments at the School of Computing and the Department of Electrical and Computer Engineering, Queen’s University. He is also the Founder and the Director of the Navigation and Instrumentation Research Lab at RMCC. He has published two books, four book chapters, and more than 270 papers in journals, magazines, and conference proceedings. His research interests include global navigation satellite systems, wireless positioning and navigation, indoor positioning, and multi-sensor fusion targeting applications related to autonomous systems, intelligent transportation, road information services, crowd management, and the vehicular Internet of Things. His research led to 13 patents and several technologies licensed to the industry in position, location, and navigation systems.
\end{IEEEbiography}

\end{document}

%% file: Preample.tex
\usepackage{cancel}
\usepackage{tabularx,booktabs}
\usepackage{lipsum}
\newcolumntype{C}{>{\centering\arraybackslash}X} 

\usepackage{cite}

\usepackage{wrapfig}
\ifCLASSINFOpdf
   \usepackage[pdftex]{graphicx}
   \graphicspath{{../pdf/}{../jpeg/}}
   \DeclareGraphicsExtensions{.pdf,.jpeg,.png}
\else
   \usepackage[dvips]{graphicx}
   \graphicspath{{../eps/}}
   \DeclareGraphicsExtensions{.eps}
\fi
\usepackage[cmex10]{amsmath}
\usepackage{textcomp,gensymb}
\usepackage{relsize}
\usepackage{color,soul}

\usepackage{scalerel} 
\usepackage{tikz} 
\usetikzlibrary{svg.path}

\definecolor{orcidlogocol}{HTML}{A6CE39}
\tikzset{
	orcidlogo/.pic={
		\fill[orcidlogocol] svg{M256,128c0,70.7-57.3,128-128,128C57.3,256,0,198.7,0,128C0,57.3,57.3,0,128,0C198.7,0,256,57.3,256,128z};
		\fill[white] svg{M86.3,186.2H70.9V79.1h15.4v48.4V186.2z}
		svg{M108.9,79.1h41.6c39.6,0,57,28.3,57,53.6c0,27.5-21.5,53.6-56.8,53.6h-41.8V79.1z M124.3,172.4h24.5c34.9,0,42.9-26.5,42.9-39.7c0-21.5-13.7-39.7-43.7-39.7h-23.7V172.4z}
		svg{M88.7,56.8c0,5.5-4.5,10.1-10.1,10.1c-5.6,0-10.1-4.6-10.1-10.1c0-5.6,4.5-10.1,10.1-10.1C84.2,46.7,88.7,51.3,88.7,56.8z};
	}
}

\newcommand{\orcidicon}[1]{\href{https://orcid.org/#1}{\mbox{\scalerel*{
				\begin{tikzpicture}[yscale=-1,transform shape]
				\pic{orcidlogo};
				\end{tikzpicture}
			}{|}}}}

\usepackage [hidelinks] {hyperref} 

\newcommand\tab[1][2.8cm]{\hspace*{#1}}
\newcommand\tabz[1][3.5cm]{\hspace*{#1}}

\usepackage{algorithmic}
\usepackage[ruled,vlined]{algorithm2e} 
\usepackage{amsmath,amssymb,amsfonts}

%% file: Authors.tex
\author{Qamar~Bader\textsuperscript{\orcidicon{0000-0002-4667-1710}}\,,~\IEEEmembership{Graduate Student Member,~IEEE}
        Sharief~Saleh\textsuperscript{\orcidicon{0000-0003-1365-417X}}\,,~\IEEEmembership{Member,~IEEE,}
        Mohamed~Elhabiby\textsuperscript{\orcidicon{0000-0002-1909-7506}}\,,~\IEEEmembership{Member,~IEEE}
        Aboelmagd~Noureldin\textsuperscript{\orcidicon{0000-0001-6614-7783}}\,,~\IEEEmembership{Senior Member,~IEEE}

\thanks{Manuscript received August XX, 2022; revised August XX, 2022; accepted August XX, 2022. This work was supported by grants from the Natural Sciences and Engineering Research Council of Canada (NSERC) under grant number: ALLRP-560898-20 and RGPIN-2020-03900. (\textit{Corresponding author: Qamar~Bader.})}

\thanks{Qamar Bader, Sharief Saleh, and Aboelmagd Noureldin are with the Department of Electrical and Computer Engineering, Queen's University, Kingston, ON K7L 3N6, Canada, and also with the Navigation and Instrumentation (NavINST) Lab, Department of Electrical and Computer Engineering, Royal Military Collage of Canada, Kingston, ON  K7K 7B4, Canada (e-mail: qamar.bader@queensu.ca; sharief.saleh@queensu.ca;  nourelda@queensu.ca).}
\thanks{Mohamed Elhabiby is with Micro Engineering Tech. Inc., and also with the Public Works Department, Ain Shams University, Cairo 11566, Egypt (e-mail: mmelhabi@ucalgary.ca).}
\thanks{Digital Object Identifier XX.XXXX/XXXX.2022.XXXXXXX}}

\markboth{IEEE TRANSACTIONS ON Automation Science and Engineering,~VOL. XX,~NO. XX,~APRIL 2023}%
{Bader \MakeLowercase{\textit{et al.}}: Enabling High-Precision 5G mmWave-Based Positioning for Autonomous Vehicles in Dense Urban Environments}



\IEEEpubid{\begin{minipage}[c]{\textwidth}\ \\[12pt]
		\tab 0018-9545 \copyright 2022 IEEE. Personal use is permitted, but republication/redistribution requires IEEE permission.\\
		\tabz See http://www.ieee.org/publications\_standards/publications/rights/index.html for more information.
	\end{minipage}}
